\documentclass[10pt]{iopart}

\usepackage[utf8]{inputenc}
\usepackage[english]{babel}
\usepackage[style=trad-abbrv,backend=biber]{biblatex} 
\usepackage[hidelinks]{hyperref}       
\usepackage{booktabs}       
\usepackage{orcidlink}
\usepackage{xcolor}

\usepackage[acronym]{glossaries}

\makenoidxglossaries

\newglossaryentry{h-a-label}{
    name={human-annotated label},
    description={
    Labels that, unlike \glspl{pseudo-label}, were manually annotated by humans or, 
    in the context of \acrshort{bci}, that required the subject to execute a pre-scripted task.
    Examples of such labels in \acrshort{bci} are the imagery class that was executed during an epoch, the stimulus that was attended during an epoch, the sleep phase, the reported mental workload, the level of drowsiness, etc.
    However, we would not consider the subject's id or the electrode names as human-annotated labels
    },
}

\newglossaryentry{pseudo-label}{
    name={pseudo label},
    description={
        Pseudo labels, unlike \glspl{h-a-label}, are automatically generated labels based on data attributes (e.g., chronological order of the time samples, spatial position of the electrodes),
        or on the meta-data associated with the recordings, (i.g., subject id, subject age, electrode names, etc.)
    },
}

\newglossaryentry{unsupervised-learning}
{
    name={unsupervised learning},
    description={Machine learning algorithms that do not use \glspl{h-a-label}
    },
}

\newglossaryentry{pretext-task}{
    name={pretext task},
    description={
        Learning task on which a network can be pre-trained. Training for pretext tasks is typically done by \gls{unsupervised-learning} algorithms
    },
}

\newglossaryentry{downstream-task}{
    name={downstream task},
    description={
        Learning task on which a network can be pre-trained. Downstream tasks are typically supervised.
        In the context of BCI, downstream tasks can be the classification of imagined concepts, responses to sensory stimuli, sleep stages, emotions, mental workload, drowsiness, seizure, etc  
    },
}

\newglossaryentry{sslg}
{
    name={self-supervised learning},
    description={Subset of the \gls{unsupervised-learning} algorithms that are trained with \glspl{pseudo-label}~\cite{jingSelfSupervisedVisualFeature2021}
    }
}

\newacronym{eeg}{EEG}{electroencephalogram}
\newacronym{bci}{BCI}{brain-computer interface}
\newacronym{erp}{ERP}{event-related potential}
\newacronym{dl}{DL}{deep learning}
\newacronym[see={[Glossary:]{sslg}}]{ssl}{SSL}{self-supervised learning}
\newacronym{gan}{GAN}{generative adversarial network}
\newacronym{pca}{PCA}{principal component analysis}
\newacronym{umap}{UMAP}{uniform manifold qpproximation and projection}
\newacronym{tsne}{t-SNE}{t-distributed stochastic neighbor embedding}
\newacronym{fft}{FFT}{fast Fourier transform}
\newacronym{fitsne}{FIt-SNE}{\gls{fft}-accelerated interpolation-based \gls{tsne}}
\newacronym{svm}{SVM}{support vector machine}
\newacronym{sae}{SAE}{sparse autoencoder}
\newacronym{mae}{MAE}{masked autoencoder}
\newacronym{dae}{DAE}{denoising autoencoder}
\newacronym{vae}{VAE}{variational autoencoder}
\newacronym{cnn}{CNN}{convolutional neural network}
\newacronym{dnn}{DNN}{deep neural Network}
\newacronym{mlp}{MLP}{multilayer perceptron}
\newacronym{ica}{ICA}{independent component analysis}
\newacronym{nlp}{NLP}{natural language processing}
\newacronym{ssvep}{SSVEP}{steady-state visually evoked potential}
\newacronym{cvep}{c-VEP}{code-modulated visually evoked potential}

\addto\extrasenglish{
   
}
\addto\extrasenglish{
   
}
\addto\extrasenglish{
   
}
\newcommand{\mysubsubsection}[1]{\subsubsection{#1.}}

\DeclareBibliographyCategory{needsurldate}
\renewbibmacro*{url+urldate}{%
    \printfield{url}%
    \iffieldundef{urlyear}%
    {}%
    {
        \ifcategory{needsurldate}%
        {%
            \setunit*{\addspace}%
            \printurldate
        }%
        {}%
    }%
}
\newcommand{\entryneedsurldate}[1]{\addtocategory{needsurldate}{#1}}

\begin{document}

\title[Review of Deep Representation Learning Techniques for BCI and Recommendations]{Review of Deep Representation Learning Techniques for Brain-Computer Interfaces and Recommendations}

\author{
 Pierre Gueschel\footnotemark[1]\,\orcidlink{0000-0002-8933-7640},
 Sara Ahmadi\footnotemark[1]\,\orcidlink{0000-0001-5049-8554},
 Michael Tangermann\,\orcidlink{0000-0001-6729-0290}
}
\footnotetext[1]{These authors contributed equally to this work.}

\address{  
    Donders Institute for Brain, Cognition and Behavior,
    Radboud University,\\
    Nijmegen, Netherlands
}

\ead{pierre.guetschel@donders.ru.nl}


\begin{abstract}

In the field of \glspl{bci}, the potential for leveraging deep learning techniques for representing \gls{eeg} signals has gained substantial interest. 
This review synthesizes empirical findings from a collection of articles using deep representation learning techniques for \gls{bci} decoding, to provide a comprehensive analysis of the current state-of-the-art. 
Each article was scrutinized based on three criteria: (1) the deep representation learning technique employed, (2) the underlying motivation for its utilization, and (3) the approaches adopted for characterizing the learned representations.
Among the 81 articles finally reviewed in depth,
our analysis reveals a predominance of 31 articles using autoencoders. 
We identified 13 studies employing \gls{ssl} techniques, among which ten were published in 2022 or later, attesting to the relative youth of the field. However, at the time being, none of these have led to standard foundation models that are picked up by the BCI community.
Likewise, only a few studies have introspected their learned representations. 
We observed that the motivation in most studies for using representation learning techniques is for solving transfer learning tasks, 
but we also found more specific motivations such as to learn robustness or invariances, as an algorithmic bridge, or finally to uncover the structure of the data. 
Given the potential of foundation models to effectively tackle these challenges, we advocate for a continued dedication to the advancement of foundation models specifically designed for \gls{eeg} signal decoding by using \gls{ssl} techniques. 
We also underline the imperative of establishing specialized benchmarks and datasets to facilitate the development and continuous improvement of such foundation models.
\end{abstract}


%
%
\submitto{\JNE}
%
\maketitle
%
\ioptwocol

\glsresetall


\section{Introduction} \label{introduction}
Representing high-dimensional data elements into lower-dimensional vectors usually facilitates their processing by subsequent machine learning algorithms. This representational process is called \textit{embedding} and is carried out by an \textit{embedding function}. The low-dimensional vectors obtained are called \textit{embedding vectors} or \textit{embeddings} also for short. 
In deep learning, we typically consider that any intermediate data representation of neural networks can be regarded as embeddings.
However, in this article, we will focus on studies that either explicitly introspect, or that use algorithms that directly optimize the embedding. 
The terms \textit{embedding vector} and \textit{representation} will be used interchangeably in the following.

For this article, we consider embeddings in the context of \glspl{bci}. \Glspl{bci} are systems that allow direct communication from a subject's brain to a computer, omitting motor output. This is realized by recording brain activity, decoding the recorded signals and interpreting the decoded information. The decoding outcomes are interpreted either as brain states of interest which are to be monitored over time, or as control commands that are send to the computer.
State-of-the-art \glspl{bci} for controlling devices or computer applications decode changes of brain activity which are a response to either an external stimulus or the result of a subject actively executing a mental task.
As brain activity is predominantly recorded via \gls{eeg}, we will primarily focus on this type of \gls{bci} in this article. This recording modality has the advantage a relatively low cost compared to other recording techniques such as magnetoencephalography and functional magnetic resonance tomography, and is often preferred over local field potentials, signals from stereotactic EEG or electrocardiography due to its non-invasiveness. \Glspl{bci} based on \gls{eeg} reflects electrical brain activity with minimal delay and a relatively high temporal resolution, allowing to build applications which impose high demands in the temporal domain.
Here, the data elements to be represented as embedding vectors consist of short windows of \gls{eeg} time-series signals, i.e., epochs or trials. 

Embedding vectors serve as a fundamental framework for \textit{transfer learning} or \textit{domain adaptation}. These terms describe the approach to employ data, hyperparameters, trained models or other information which had been obtained from earlier recordings or earlier users to a novel recording or user. 
Transfer learning is of particular interest within the BCI community~\cite{jayaramTransferLearningBraincomputer2016,wuTransferLearningEEGBased2022}, as it may help to solve a central problem in research and clinical applications of BCI: Training a decoding method from scratch for a novel user or session is challenged by a lack of time. 
%
However, there are several caveats to consider. 
Firstly, there may be potential changes in strategies between subjects. 
Secondly, the exact location, timing, intensities, and frequencies of neural activity may vary between individuals, along with their brain morphologies. 
Thirdly, individual lesions in stroke patients and individual progress patterns and deficits in neurodegenerative diseases can affect transfer learning.
Lastly, various confounding factors such as medication, sleep patterns, imprecise electrode placements, artefacts and environmental factors can also lead to non-stationarities in the signals.
%

This review is motivated by observing a growing number of publications in the  \gls{bci} field during recent years, which have used embedding techniques.
However, it is unclear which techniques are most commonly used to learn embeddings for BCI. In addition, it is not established which alternative deep learning approaches so far have remained unexplored in BCI and which benefits they could bring. Finally, it might also be valuable for the BCI community to see which purposes exactly serve the different types of embedding and how they can be benchmarked and introspected.

In this review article, our focus is three-fold.
Firstly, we focus on the potential motivations researchers have for using embeddings.
Indeed, we observe that the use of DL-based representations in BCI can be motivated by multiple different reasons. 
%
Secondly, this review article aims to draw the spectrum of possible methods that
can be used for embedding learning, and more generally feature learning,
using \gls{dl} for \gls{bci}
applications. 
Lastly, we look at introspection techniques for embeddings.
These techniques can be informative for comparing and evaluating embeddings, and reveal what type of information can be obtained from them.
Throughout this article we will explore the literature on deep representation learning for BCI. 
Additionally, we will have a view on studies involving non-BCI EEG data with the purpose to potentially identify research gaps in the BCI field. 
Furthermore, we will also provide examples from leading deep learning domains, such as computer vision, \gls{nlp}, and speech processing, to illustrate the potential of deep representation learning for the field of BCI.

This article is organized as follows: in \autoref{methodology}, our approaches for retrieving and filtering articles and for extracting information are introduced.
\autoref{motivation}
explores different possible motivations for learning an
embedding or using an algorithm that intrinsically does that.
In \autoref{algorithms}, we will draw a list of the different
approaches that have been used to learn embeddings in the BCI field. We will also
report algorithms used in neighbouring fields for non-BCI, but EEG data, as they could be
relevant for BCI.
Finally, in \autoref{introspection}, we will explore the
different methods that have been proposed to benchmark, qualify, and
compare the embeddings learned.

\section{Our Methodology} \label{methodology}
With the focus of this review article being on the intersection between the notions of \textit{deep representation learning} and \textit{\gls{bci}}, the first step has been to collect articles dealing with both topics simultaneously. Unfortunately, only few articles contain these exact keywords in their title or abstract and too many contain them in the main text body. Therefore, we had to establish a finer search strategy. 

This strategy consisted of first creating, for both of the two notions, a list of terms that were either equivalent or implying it. 
For example, the notion of \textit{\gls{bci}} can either be replaced by equivalents such as \textit{brain-computer interface}, and \textit{brain-machine interface}, or by specific paradigms that may imply a \gls{bci}, such as \textit{motor imagery} or \textit{event-related potential}.
Similarly, the notion of \textit{deep representation learning} can either be replaced by equivalents like \textit{deep learning} + \textit{embedding}, or by techniques that inherently obtain a deep learning-based representation, such
as \textit{autoencoders}. We noted the importance of accepting many different terms that describe potential deep learning methods which can learn an embedding. To compile this list of method terms, we used the review conducted by Roy and colleagues~\cite{royDeepLearningbasedElectroencephalography2019}.
The query finally used in search engines was the conjunction (AND) between the disjunction (OR) of all the terms of the \textit{BCI} list and the disjunction of all the \textit{deep representation learning terms}.
This query contains 65 terms in total. Note that a term can eventually contain multiple words (such as \textit{deep learning}).  AND, OR and the parenthesis are not considered terms. We restricted our search to articles published after 2014 because we did not expect any earlier relevant work involving both deep learning and BCI. Finally, we restricted our search to the titles only in the search engines.

We initially started by using three search engines: Web of Science, PubMed and Google Scholar. Unfortunately, Google Scholar did not allow for nesting terms in parenthesis. In a second attempt, we computed the disjunctive normal form of our expression (which removes the need for parenthesis) but it resulted in a 8568 terms expression which hit the 150-word limit of Google Scholar. Therefore, we had to drop Google Scholar and only used Web of Science and PubMed. The matching articles were gathered using Publish or Perish~\cite{harzing-PublishPerish2007} and organized using Zotero~\cite{vanheckeZotero2008}.

\begin{figure*}
	\centering
	\includegraphics[scale=.8]{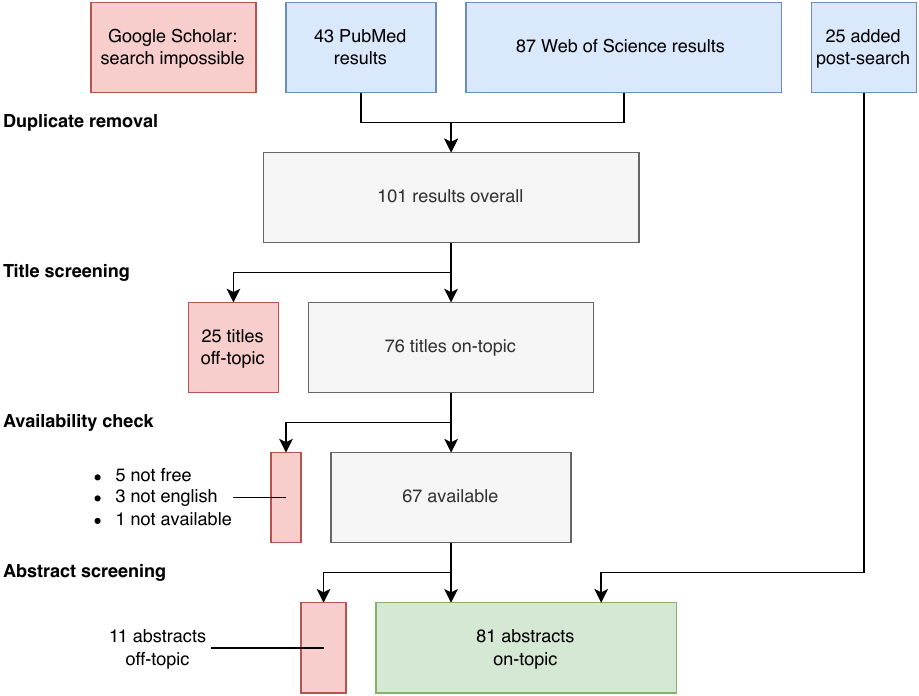}
	\caption[Articles selection process]{Flow diagram summarizing the process for selecting which articles from the initial search results to consider in this review. Blue boxes indicate the initial available sources and number of articles, red boxes indicate articles which were not considered for various reasons (see main text), and the green box represents the finally considered articles.
    }
	\label{selection_diag}
\end{figure*}

We found 87 articles from Web of Science and 43 from PubMed, which resulted in 101 articles after removing duplicates. This search was conducted on April 1\textsuperscript{st} 2024.
We read all the titles and found that 25 were off-topic, which left us with 76.
Among those, five were behind a paywall for which we did not have access, three were not in English and one was not available. This left us with 67 articles. We read the 67 abstracts and found that eleven articles were still off-topic, which left us with a final selection of 56 articles. 
We included 25 additional articles post-search directly to the final selection. They were detected either in reference sections of the 56 articles or were selected based on our prior knowledge of the field. These additional articles also include non-BCI \gls{eeg} studies which have applied techniques that can be interesting to the BCI community.
A flow diagram summarizing the selection process is provided by \autoref{selection_diag}.

 Among the resulting  56+25 articles, many use similar techniques, in particular, 34 articles employed autoencoders to learn a representation. Because of such redundant approaches we refrained from discussing every single paper, but maintained all of them in our literature list. 

While reading the resulting 81 articles, our focus was on three aspects: 
\begin{enumerate}
    \item \label{aspect:motivation} the motivation(s) for learning an embedding,
    \item \label{aspect:algorithms} the algorithms and approaches used for learning the embedding,
    \item \label{aspect:introspection} and the methods used for characterizing and introspecting the obtained embeddings.
\end{enumerate}
Our findings on these three aspects are respectively reported in \autoref{motivation}, \autoref{algorithms} and \autoref{introspection}. The first aspect, the motivation authors had to learn an embedding, was not always explicit. We estimated their motivation by reading the Introduction and Discussion sections and by identifying the problems being addressed in the article. 
However, we refrained from interpreting why the authors used a particular method or made a specific design choice.
The second and third aspects, algorithms and characterisation/introspection techniques, were usually clearer and less prone to interpretation. Their identification was done by respectively reading the Methods and Results sections.

Because the goal of any review, including this one, is to provide an overview of the state-of-the-art for a research topic, the aspect~(\ref{aspect:algorithms}) on algorithms is central. It allows researchers entering the field to choose from the complete panel of methods at their disposition. 
However, to choose between the algorithms presented, researchers need to understand which needs each of these algorithm addresses, i.e., for what reason should one algorithm be preferred over another? Hence, the aspect~(\ref{aspect:motivation}) on motivations. Moreover, this section can also be used to help researchers entering the field identify their own motivations by reviewing a list of potential ones.
Finally, the aspect~(\ref{aspect:introspection}) on introspection techniques is necessary because the algorithms presented in this review are quite specific and the introspection methods commonly used in BCI to characterise classical machine learning algorithms may be of limited use only in this BCI context. Additionally, most of the introspection techniques presented simply take embedding vectors as input, without making assumptions about the algorithm that had been used to learn the embedding. Therefore, we will see how certain introspection methods commonly paired with certain algorithms can actually be paired with many other algorithms described in \autoref{algorithms}.

\section{Motivations to learn embeddings} \label{motivation}

In this section, we will report on which motivations were identified as leading to the use of a DL-based embedding or to the use of a method that inherently learns one.
Please note that the motivations we list in the following are not exclusive and that the authors of a study can have multiple motivations for learning an embedding.
Many of the motivations listed in this section are special cases of \emph{transfer learning}, which turned out to be a motivation in the large majority of the articles we reviewed for learning an embedding. 
In the following subsections, we  explain why embeddings seem specifically suited for transfer learning in BCI. 

\subsection{Improve classification accuracy}

Undoubtedly, most articles share the objective of improving the classification accuracy over the state of the art in a particular scenario. 
However, in some articles the reference to embeddings is only motivated by that reason (e.g.~\cite{doseEndtoendDeepLearning2018,liuDistinguishableSpatialspectralFeature2021}).
While this is a legitimate motivation by itself, it does not tell us much about how the embedding is learned or why a particular algorithm was used. For this reason, we will not elaborate more on this motivation.


\subsection{Learning to become robust to noise}\label{robust}
EEG is sensitive to many sources in addition to the signal of interest.
These additional sources can be, for example, non-physiological noise picked up by the system, muscular artefacts or other non-neural biosignals, or background brain activity.
These additional sources are considered, in many cases, not relevant to the BCI task and typically they do not help the decoding, as they may fluctuate over time, showing non-stationary distributions.
We observe that the least restrictive experimental protocols tend to be  most affected by these undesired sources. Examples are dry or water-based EEG systems that are easier and faster to set up than gel-based ones, but are also more prone to noise.
Similarly, real-world conditions are less constrained than lab conditions where subjects are asked not to blink and to remain still during the recordings. Here, the former will lead to more artefacts and non-stationarities in the signals.
Therefore, it is desirable to have systems which have learned to be robust to these additional sources.

We will see two methods that allow learning robustness.
The first one, for robustness to noise specifically, is \emph{denoising autoencoders}~\cite{chenDenoisingAutoencoderbasedFeature2021,qiuDenoisingSparseAutoencoderbased2018} that will be explained in \autoref{dae}.
The second method is through \emph{data augmentations}~\cite{rommelCADDAClasswiseAutomatic2022} which will be explained in \autoref{augmentation}.
%

\subsection{Learning invariances}
This motivation addresses experimental scenarios or protocols which record brain signals under multiple \emph{conditions}. Typical conditions are the subject id, the session number, or the source dataset. Data collected over conditions can be expected to follow different distributions.
The conditions usually are orthogonal to the main BCI classes, i.e., there can be examples with any combination of condition and BCI class.
A model is said to be invariant with respect to a condition if the representation it produces does not depend on that condition.
Learning domain-invariant representations is a typical approach to \emph{transfer learning} because it allows using the same representation on multiple data distributions.
%
The notion of invariance is close to that of robustness described in \autoref{robust}. However, we made a distinction between the two as in the case of robustness, we will systematically refer to factors that are only obstacles to the decoding (such as noise and artefacts), whereas in the case of invariance, we will refer to contextual information for which a conscious choice is made to maintain invariance.

We will see in \autoref{domain_invariant} that invariant representations can be learned through an adversarial objective~\cite{jeonMutualInformationDrivenSubjectInvariant2021} or by using deep metric learning~\cite{GuePapTan21}, see~\autoref{metric}.


\subsection{Learning from small datasets}\label{little}
The ability to learn from as little as possible data  while still reaching satisfactory classification scores is desirable in BCI systems for two main reasons:
First, it allows for the quick start of BCI applications because only a small amount of calibration data needs to be recorded.
Second, BCI datasets are quite small in general.

\emph{Shortening calibration times} can be addressed with two different types of algorithms:
algorithms able to exploit very well the few available examples of the ongoing session~\cite{sosulskiUMMUnsupervisedMeandifference2023,thielenFullCalibrationZero2021},
or algorithms that can take advantage of existing data recorded before the current session such that only little adaptation is needed for the ongoing session~\cite{GuePapTan22,koblerSPDDomainspecificBatch2022}.
The former is generally not a strong point of deep learning models
but rather of classical machine learning models that exploit expert knowledge, e.g., in the form of domain-specific regularization approaches~\cite{sosulskiImprovingCovarianceMatrices2021,sosulskiIntroducingBlockToeplitzCovariance2022,sosulskiUMMUnsupervisedMeandifference2023,thielenFullCalibrationZero2021}.
The latter is better known as \emph{transfer learning}, and using it is nearly always motivated by the reduction of calibration times.

\emph{Tackling small datasets}  essentially involves the same principle as \textit{shortening calibration times}, hence it can also be done in two different ways: either by simply exploiting well the small existing datasets or by pre-training models on other types of datasets.

Most of the algorithms we will mention in this article allow for some form of transfer learning.
However, we will see in  \autoref{unsupervised} that unsupervised algorithms are particularly useful for this purpose as they allow the use of non-BCI EEG datasets that are much more abundant~\cite{obeidTempleUniversityHospital2016}. The unsupervised algorithms will be described in detail in Subsections~\ref{autoencoder}, \ref{ssl}, \ref{gan} and~\ref{metric}.



\subsection{Bridging heterogeneous components}


Raw data elements generally have intrinsic structures: An image has a width and a height, a text has a certain number of words, and an EEG recording has a duration, sampling rate and a specific spatial layout of recording channels.
These intrinsic structures define a relation of the features contained in an example: the pixels of an image, the words of a text, and the samples of an EEG recording follow specific orders.
Each data type requires specialized layers in DL architectures and/or pre-processing steps to capture their internal structures.
These layers and steps are necessary to transform the data elements into \textit{forms} that allow for better processing by the following classifiers (or classification layers).
As explained in the Introduction (\autoref{introduction}), these forms are called embeddings, i. e., collections of features, where the eventual purpose of each feature is automatically defined by the training algorithm.
We observed in the literature that an embedding can be used as an algorithmic bridge in multiple ways:

\mysubsubsection{Between different types of algorithms}
Because there is no a priori hypothesis about the features of an embedding, virtually any classification or regression algorithm can use embedding features as input. Therefore, it is common to use an embedding as a bridge between different types of decoding algorithms~\cite{GuePapTan22,xuRepresentationLearningMotor2021}.
A typical scenario uses a deep learning model as feature extractor and a classical machine learning model to decode those features.
This configuration can be employed for transfer learning scenarios.

\mysubsubsection{Between different data types}
Because the purpose of each feature in an embedding is learned automatically by the algorithm, it is possible to design the learning task such that different data types can be projected into a common embedding space.
Even if every data type requires a different processing pipeline, they can all produce an embedding of the same dimensionality. Then, techniques exist to align the embedding spaces of different data types according to their semantic similarity.
This so-called \textit{joint embedding learning} is commonly used to relate image and text data into an embedding~\cite{kirosUnifyingVisualSemanticEmbeddings2014,liVisualBERTSimplePerformant2019}, but first publications have now shown, how joint embeddings can be learned also for EEG and MRI data~\cite{ferriStackedAutoencodersNew2021}.

\mysubsubsection{Between different recording systems}
Despite existing norms for EEG electrode placement, there are many different EEG systems available, all with particularities and slight variations. Additionally, not all datasets are recorded using the same set or even the same number of channels. 
To address this obstacle, recent studies have begun investigating architectures that can receive recordings from multiple different channel sets as input \cite{yaoEmotionClassificationBased2024,guetschelSJEPASeamlessCrossdataset2024,wangNovelAlgorithmicStructure2023,yangBIOTCrossdataBiosignal2023,bruschMultiviewSelfsupervisedLearning2023}, see \autoref{mask-ssl}. These architectures show promise for transfer learning or for handling corrupt channels.

\mysubsubsection{To enforce multiple objectives}
Finally, the embedding can enable the combination of multiple high-level objectives as in the following example:
During training (in opposition to joint embeddings) a single processing pipeline is employed to create the embedding layer. From this point on the processing can be split into multiple branches, each computing a specific objective. Finally, the global loss would be a weighted sum of these different objectives.
In this scenario, the embedding provides a high-level representation of the data which tends to satisfy all the different objectives.
One could want to enforce multiple objectives simultaneously, for example, to both optimize the performance of the model on a BCI task and to obtain a subject-independent representation as proposed by Ozdenizci and colleagues~\cite{ozdenizciLearningInvariantRepresentations2020}, see \autoref{domain_invariant}, 
or to simultaneously enforce a supervised and an unsupervised objective for the purpose of mitigating the risk of overfitting and enhancing generalization, see Li and colleagues~\cite{liMultitaskCollaborativeNetwork2024} and \autoref{ssl}. 


\subsection{Uncover the structure of the data}
Finally, \acrfull{ssl} algorithms differ from the aforementioned approaches by not relying on labelled data (see \autoref{ssl}). Instead, they learn representations in a data-driven manner. As a result, the patterns that emerge when summarizing datasets using visualization methods such as \acrshort{umap} (see \autoref{projection}) may reveal underlying structures in the data, rather than simply reflecting the prior assumptions of an experimenter. In this direction, Banville and colleagues demonstrated that the representations learned through SSL contained structures that translated physiological and clinical phenomena~\cite{banvilleUncoveringStructureClinical2021}. This finding highlights the potential of SSL algorithms to uncover meaningful structures in complex data.



\section{Approaches used to obtain embeddings} \label{algorithms}

Here, we will try to draw an exhaustive list of all the algorithms and methods that can be put into place to meet the objectives described in \autoref{motivation}. We distinguish two types of algorithms which learn representations in either a supervised or unsupervised way.



\mysubsubsection{Supervised methods} \label{supervised}
We say an algorithm is supervised when it directly exploits examples with \glspl{h-a-label}. In BCI, such labels can for example be the type of mental imagery task being executed, or the stimulus attended during an epoch, i.e., the BCI classes. In general, such examples have to be recorded under controlled conditions where the participant has to execute a pre-scripted task, as opposed to online/free BCI control.
The representations learned in this way are typically only tailored for the task corresponding to the labels and do not generalize well.

Not many of the reviewed articles learn a representation in this supervised manner. Nevertheless, in \autoref{hidden_layer}, we described one very simple example. In \autoref{metric}, we will see how metric learning can also be used to learn embeddings in a supervised way. 
Yet, supervised learning is often used to fine-tune pre-trained models. In such a case, the supervised algorithm does not properly learn the embedding but rather uses it and eventually improves it.

\mysubsubsection{Unsupervised methods} \label{unsupervised}
Algorithms implementing \gls{unsupervised-learning} do not use \glspl{h-a-label}~\cite{jingSelfSupervisedVisualFeature2021}.
This bears the advantage of being able to be trained on any raw EEG signal, i.e., not necessarily recorded under a BCI protocol,
which is far more abundant than labelled recordings of BCI sessions.
It is generally the class of unsupervised algorithms that is employed to train so-called \textit{foundation models}, i.e., general-purpose embeddings trained from large amounts of data.
In a typical scenario, an unsupervised algorithm pre-trains a neural network on what is called a \gls{pretext-task},
and a supervised algorithm later on fine-tunes it, or part of it, on what is called a \gls{downstream-task}.
In the context of BCI, downstream tasks can be the classification of an imagined or executed movement, of attended vs.~ignored stimuli, of sleep stages, the detection of seizures, emotions, or the regression of the level of mental workload, of the level of drowsiness, etc. We will see in the following sections some pretext tasks that can or could be used in BCI.

The overall goal is typically to obtain a good score on the downstream task,
and the goal of the pretext task is to provide a pre-trained network
that can already extract information relevant to the downstream task.
If the pretext task manages to learn information relevant to the downstream task,
the workload of the latter is alleviated.
This way, the network might need fewer examples  from the downstream task to be fine-tuned.
Because the pretext and downstream tasks are inevitably different to some extent,
the data features that must be learned to solve each of them are also different.
This leads to the question of how similar a particular pretext task is compared to actual BCI tasks (i.e., downstream tasks), or slightly rephrased, whether a given pretext task is general enough to learn a representation that is sufficiently general to contain information that is relevant to (various) BCI classification tasks.

Many different unsupervised algorithms exist, each with its own way of relating to the notion of an embedding. 
In \autoref{autoencoder}, we will review articles realizing autoencoder paradigms and variants of it as pretext tasks.
Then, in \autoref{ssl}, we will look at the subfield of unsupervised learning that uses \glspl{pseudo-label}, called \acrfull{ssl}.
In \autoref{gan}, we will see how \acrfull{gan} can be used for unsupervised representation learning. 
Finally, in \autoref{metric}, we will show how metric learning techniques can be realized in unsupervised manners.

\subsection{Hidden layer of a classification network} \label{hidden_layer}
The simplest approach one can think of to obtain an embedding is to train a feed-forward neural network with at least one hidden layer on a classification task, i.e., using a cross-entropy loss.
Here each output node of the network will correspond to a specific class.
Receiving a novel input example, the task of each output node is to predict the probability that the true class label is that of the node in question.
To obtain as many outputs as there are classes, there is usually a fully connected layer that linearly maps the latent representation from the last hidden layer to these outputs.
This last latent representation can be considered as an embedding, and it is common to treat it as such~\cite{GuePapTan22}, i.e., re-using that representation for other purposes, analysing the information contained in this latent representation, etc.
This method has the advantage of being extremely simple to implement.
In addition, it is generally more computationally efficient than most unsupervised methods.
However, this simple method produces representations that are very targeted, which means they are typically not transferable to other tasks and sometimes not to other distributions as well.

To obtain more flexible representations, it is common to use multiple objectives simultaneously, i.e., multi-task learning.
For example, one can enforce the representation to be invariant to the subject in addition to allowing the classification of mental imagery tasks. We will see in detail how such domain-invariant representations can be learned in \autoref{domain_invariant}.
Another approach consists of representations that allow the reconstruction of the original input.
This can be achieved by combining the classifier with an autoencoder~\cite{ditthapronUniversalJointFeature2019}, c.f. \autoref{autoencoder}.

\subsection{Autoencoder} \label{autoencoder}
Autoencoders encompass relatively well-established methods for unsupervised representation learning~\cite{bankAutoencoders2021}.
In their canonical form, their architecture can be decomposed into two parts: an \textit{encoder} and a \textit{decoder}. The encoder takes as input an example \(X\) and returns a hidden representation \(z\). The decoder takes as input the representation \(z\) and returns an estimation of the original input \(\hat{X}\). The objective is to minimize the difference between the estimated and the original inputs \(\hat{X}\) and \(X\).
Typically, the dimension of the hidden representation \(z\) is smaller than that of the input \(X\) such that the network learns to extract important features from the input data, which are necessary to reconstruct the original input.
Autoencoders are unsupervised and can virtually be applied to any type of input.
Over the years, a significant amount of variants were developed. This is reflected in the 34 studies we have found using autoencoders.

\mysubsubsection{VAE}\label{vae}
\Glspl{vae}
is a variant in which the latent representation predicted by the encoder is not directly fed to the decoder, but rather used as parameters for a random distribution, typically an isotropic Gaussian.
Then, a point sampled from this distribution is given to the decoder as input~\cite{bankAutoencoders2021}.
Modelling the problem in this way allows to gain control over the meaning of a latent representation, which would not be possible with a regular autoencoder.
Additionally, depending on the distribution used, it can force the hidden representation to have disentangled features~\cite{higginsEarlyVisualConcept2016}.
These reasons might explain why \glspl{vae} have been well adopted by the BCI community.
For example, Ozdenizci et al.~used them in combination with adversarial networks (c.f. \autoref{adversarial}) to learn subject-invariant representations of motor imagery recordings~\cite{ozdenizciTransferLearningBrainComputer2019}.

\mysubsubsection{DAE and MAE}\label{dae}\label{mae}

\Glspl{dae} and \glspl{mae}
are variants where the input for the encoder is corrupted, either by adding noise or by masking parts of it.
Nevertheless, the output of the decoder is still compared to the original unaltered input.
This way, the network can learn how to handle noisy or corrupted input examples.
These are interesting characteristics for EEG data and explain why they are commonly used~\cite{chienMAEEGMaskedAutoencoder2022,xieMotorImageryEEG2020, qiuDenoisingSparseAutoencoderbased2018, chenDenoisingAutoencoderbasedFeature2021}. 

\mysubsubsection{SAE}\label{sae}
\Glspl{sae}
differ from regular autoencoders in that they enforce sparsity in the hidden representation through an additional loss term. 
Sparsity means that only a small number of dimensions of the representation are simultaneously non-zero.
The sparsity is typically enforced by adding a penalty term which enforces a minimal Kullback-Leibler (KL) divergence~\cite{kullbackInformationSufficiency1951} between the average activation of each hidden unit and a sparsity hyperparameter.
Liu et~al.~\cite{liuEEGBasedEmotionClassification2020} proposed a deep learning architecture for EEG-based emotion recognition, which consists of a \gls{cnn} for feature extraction, followed by a \gls{sae} and a \gls{mlp} classifier. 
Their approach involves training the \gls{cnn} in a supervised manner, then discarding the linear classification layer and training the \gls{sae} in an unsupervised way using the \gls{cnn} features. 
Finally, the \gls{mlp} is trained in a supervised way using the \gls{sae} features. 
Their results support that the proposed approach outperforms other methods, including the \gls{cnn} part alone.
Qui et~al.~\cite{qiuDenoisingSparseAutoencoderbased2018} proposed an approach combining denoising and sparse autoencoders for seizure detection. They demonstrated that the joint use of these two techniques resulted in better performance than either approach used independently. The \gls{dae} component was shown to prevent overfitting and improve the robustness of the model to noise, while the \gls{sae} allowed for the learning of higher-level features in the EEG data. 


\subsection{Self-Supervised Learning (SSL)} \label{ssl}
\Gls{ssl} is a form of \gls{unsupervised-learning} that uses \glspl{pseudo-label}, or automatically generated labels, to train a network~\cite{jingSelfSupervisedVisualFeature2021}.
Designing these \glspl{pseudo-label} is typically done by using known properties of the data, e.g., an example is similar to itself, a time sample comes before the next one, samples from different channels are correlated, etc.
A pseudo label could be, for example, the chronological order of different time windows of a recording.
The unsupervised learning is guided by the necessity to create pretext tasks which either are similar to the downstream tasks or sufficiently general to learn features that would be relevant for them.
\gls{ssl} have been extremely successful in the fields of
computer vision~\cite{chenSimpleFrameworkContrastive2020,bardesVICRegVarianceInvarianceCovarianceRegularization2022},
speech processing~\cite{baevskiWav2vecFrameworkSelfSupervised2020}
and \gls{nlp}~\cite{devlinBERTPretrainingDeep2019}. 

As the topic is fairly new, it is still relatively unknown within the BCI community.
Out of the 13 studies we found using SSL with BCI data, 10 were published after 2021.

\mysubsubsection{SSL Tasks Using Temporal Structure}
In their early 2021 study, Banville et al.~\cite{banvilleUncoveringStructureClinical2021} compared three different SSL pretext tasks that all exploit the temporal structure of EEG recordings.
The first task, \textit{relative positioning}, consists of predicting whether two time windows were within a certain distance or further apart.
The second task, \textit{temporal shuffling}, requires to determine if three time windows are in the correct order or not.
The last task, \textit{contrastive predictive coding}, uses a number of consecutive windows and distractors that had been sampled elsewhere. The first two consecutive windows constitute the context. The task is to predict, for each other window, whether it is following the context or is a distractor.
A similar approach was taken by Ou and colleagues~\cite{ouImprovedSelfsupervisedLearning2022}.

\mysubsubsection{Masking-based SSL}\label{mask-ssl}
 Also in 2021, Kostas et al.~\cite{kostasBENDRUsingTransformers2021} published an SSL method called wav2vec that was originally developed for speech processing~\cite{baevskiWav2vecFrameworkSelfSupervised2020}.
Wav2vec shares similarities with masked autoencoders (c.f.~\autoref{mae}), where temporal regions of the signal are masked, but directly optimises the distance between embedding vectors instead of going back to the input space.
This work sparked a number of studies, which also experimented with temporal masking strategies~\cite{heSelfsupervisedLearningBased2022,bruschMultiviewSelfsupervisedLearning2023,cuiNeuroGPTDevelopingFoundation2023,liMultitaskCollaborativeNetwork2024,foumaniEEG2RepEnhancingSelfsupervised2024}.

Most masking-based SSL studies rely on transformer architectures~\cite{vaswaniAttentionAllYou2017}. This is because the attention mechanism of transformers allows the network by design to selectively focus on the unmasked portions of the signal and predict the masked portions.
Yang and colleagues pioneered the use of transformers with independent encoding of the EEG channels~\cite{yangBIOTCrossdataBiosignal2023}. This novel approach enables the exploration of masking strategies over the channels that allow to pre-train an attention mechanism over the channel structure. Such pre-trainings pave the way for the development of pre-trained models that are independent of specific channel sets. 
Subsequent studies by Li~et~al.~and Guetschel~et~al.~further expanded on this concept with a domain-inspired spatial masking strategy~\cite{liMultitaskCollaborativeNetwork2024,guetschelSJEPASeamlessCrossdataset2024}.



\mysubsubsection{Augmentation-based contrastive SSL}\label{augmentation}
As a third pioneering article (2020), Mohsenvand and colleagues experimented with SSL based on data augmentations~\cite{mohsenvandContrastiveRepresentationLearning2020}, and were followed by Yand and colleagues~\cite{yangSelfsupervisedEEGRepresentation2023}.
The general idea of such methods is to  \textit{first} sample two data augmentations from a pre-defined family of plausible augmentations.
In the image domain such augmentations could be a combination of cropping,  rotation, colour shift and addition of noise.
Sampling an augmentation would then mean sampling a set of parameters for cropping, rotating, shifting the colours and adding noise.
\textit{Second}, the two augmentations are applied to the examples of the batch, resulting in two augmented versions of every example.
\textit{Third}, the augmented examples are passed to a network to obtain embeddings.
\textit{Finally}, the loss function enforces the embeddings of the two versions of each example to be either similar (non-contrastive) or to share a higher similarity with each other than with the representations of the other examples in the batch (contrastive).

Plausible and efficient augmentations are different for each domain.
Domain knowledge about EEG signals and neural processes can guide the design of novel, plausible data augmentations for BCI. 
For example, a slight shift in the orientation of the source dipoles~\cite{zlatovPhysiologyinformedDataAugmentation2022} or of their amplitude~\cite{castano-candamilPosthocLabelingArbitrary2019} are plausible and can be used as data augmentation.
However, it is not established which augmentations would be the most efficient for self-supervised representation learning with BCI data and EEG processing in general.
Rommel et~al.~\cite{rommelDataAugmentationLearning2022} reviewed data augmentations which have already been tested on EEG data.
The authors underline that different BCI tasks require different data augmentations.
They also recognize that the list of augmentations that have already been introduced for or tested on EEG signals is probably not exhaustive and new ones could still be discovered.
%

Overall, the field of \gls{ssl} remains relatively unexplored in BCI but is part of the \textit{hot topics} in deep learning. The BCI community would probably gain much from exploring \gls{ssl} further.

\subsection{Adversarial network-based training} \label{adversarial}

Some learning objectives are complex, i.e., non-trivial to compute.
In other words, they can not simply be evaluated by non-parameterised loss functions such as the mean squared error, or the cross-entropy loss.
Such complex objectives may, for example, maximize the level of realism of a generated example~\cite{fahimiGenerativeAdversarialNetworksbased2021}, minimize the mutual information between two embeddings~\cite{jeonMutualInformationDrivenSubjectInvariant2021}, or minimize the amount of domain-specific information present in a latent representation~\cite{ozdenizciTransferLearningBrainComputer2019}.
To enforce such complex objectives, it is possible to utilize \textit{auxiliary} neural networks.
These auxiliary networks can deliver richer feedback to the main networks than a non-parameterized loss functions.
The specificity of the methods presented in this section is that their auxiliary networks all learn to maximize their objective, whereas their main networks are trained to minimize it.
For this reason, these auxiliary networks are actually called \emph{adversarial networks}.
In mathematical terms, the objective function being optimized is the \textit{minimum} over the possible main networks of the \textit{maximum} over the possible adversary networks of the loss function.

\mysubsubsection{Generative adversarial networks}\label{gan}
A~specific case an adversarial network training can be found in \glspl{gan}, where the main network is called \textit{generator} and the adversary network is called \textit{discriminator}.
The generators are optimised for generating realistic data from random noise vectors,
and the discriminators are trained to discriminate if the examples they receive as input are real or have been artificially created by the generators~\cite{goodfellowGenerativeAdversarialNetworks2020}.

In the image domain, Radford et al.~emphasized that \gls{gan}s can be used
to learn representations in an unsupervised way~\cite{radfordUnsupervisedRepresentationLearning2016}.
Indeed, the discriminators can take as input any kind of data
and are trained on a relatively high-level task
(depending on how good the outputs of the generator are),
such that  there is a good chance to find features relevant to other learning tasks in their hidden representations.
Furthermore, it has been shown by Vondrick and colleagues that relevant features can be learned from video data this way~\cite{vondrickGeneratingVideosScene2016}.
However, \gls{gan}s are known to be difficult to train (long and unstable) 
such that their use as unsupervised feature extractors remains marginal in general.

In the context of BCIs, we did not find any article using them for that purpose, but it could be worth investigating this further.
However, there are many successful examples of \gls{gan}s being used to generate fake BCI examples in the context of data augmentation~\cite{hartmannEEGGANGenerativeAdversarial2018,gaoGenerativeAdversarialNetwork2022,fahimiGenerativeAdversarialNetworksbased2021}.

\mysubsubsection{Domain-invariant representations} \label{domain_invariant}
A~particular use of adversarial networks is to learn domain-invariant representations.
In the context of BCI, the domains are typically the subjects or the sessions, but they can also be the recording equipment, the dataset, the stimulation parameters, or factors.
A domain-invariant representation is particularly interesting for cross-domain transfer, e.g., if we have a decoding model pre-trained on multiple subjects and want to apply it to a novel one.
Transfer learning is one of the current challenges of BCI~\cite{wei2021BEETLCompetition2022}.

To enforce representations to be domain-invariant, a usual objective of adversarial networks is to identify from which domain the input examples come.
In this context, the auxiliary networks taking the adversary role are often called \textit{domain discriminators}.
The objective of the main network is partly to fool the domain discriminator by leaving no domain-specific information in the representations they generate and partly to complete another task such as classification~\cite{ganinUnsupervisedDomainAdaptation2015,serdyukInvariantRepresentationsNoisy2016}.

Özdenizci and colleagues implemented this for subject-independent motor imagery feature extraction~\cite{ozdenizciTransferLearningBrainComputer2019, ozdenizciLearningInvariantRepresentations2020}. 
Their adversarial network is trained to discriminate the subjects and is paired with a \gls{vae} (see \autoref{vae}).

Jeon and colleagues argued that using a domain discriminator can introduce problems like discard class-relevant information, also referred to as \textit{negative transfer~\cite{panSurveyTransferLearning2010}}.
They instead trained their adversarial network to estimate the mutual information between  \textit{class-relevant} and \textit{class-irrelevant features}~\cite{jeonMutualInformationDrivenSubjectInvariant2021}. Their main network's objectives are to minimize the mutual information between the \textit{class-relevant} and \textit{class-irrelevant features} and to classify motor imagery examples using the \textit{class-relevant features}. Note that Jeon et~al. used the term \textit{adversarial learning} which to our knowledge is slightly unconventional in this context while it commonly describes attacks on models (i.e., reverse engineering, trying to fool the models, etc.).

\subsection{Deep metric learning} \label{metric}
Deep metric learning is a sub-field of deep learning focussed on training neural networks to embed examples into vector spaces whose metrics implement \textit{notions of similarity} between examples.
A typical notion of similarity would be class membership: examples from the same class would be considered more similar, i.e., close to each other in the embedding space, than examples from different classes.
It is possible to define notions of similarity even without \glspl{h-a-label} by using \glspl{pseudo-label} or by simply considering that an example is \textit{similar} to itself. These cases will also be referred to as \gls{ssl} (see \autoref{ssl}).

A loss function commonly used for deep metric learning is the \textit{triplet loss}~\cite{schroffFaceNetUnifiedEmbedding2015}.
The triplet loss takes three examples as input: an anchor, a positive and a negative example. The anchor and the positive examples are expected to be similar, while the anchor and the negative ones are dissimilar.
The loss minimizes the distance between the representations of anchor and positive and at the same time also maximizes the distance between the representations of anchor and negative.
Most of the loss functions used for deep metric learning are relatively similar to the triplet loss or the \textit{contrastive loss}~\cite{hadsellDimensionalityReductionLearning2006} which does not involve negative examples.
A particularity of deep metric learning techniques is, that even if they can be supervised,
they directly optimize the representations or the embedding space. In other words, the data representations are obtained explicitly and not merely as a side-product.
Deep metric learning is frequently used in computer vision for tasks such as face recognition~\cite{schroffFaceNetUnifiedEmbedding2015}  or place recognition~\cite{arandjelovicNetVLADCNNArchitecture2016}.
In these tasks, the class is the identity of the person on the picture or the place where the photo was taken.
%
These tasks have in common that they usually have only a few examples per class but many different classes.
Additionally, they require models which can work on classes that were unseen during training.
This last requirement makes it impossible to use regular classifiers for these tasks.

Schneider et al.~\cite{schneiderLearnableLatentEmbeddings2022} used a variation of the triplet loss and introduced a novel triplet sampling scheme for learning embeddings for neural data jointly with behavioural data and/or time. Triplet sampling holds an important role in metric learning \cite{schroffFaceNetUnifiedEmbedding2015}. While the authors tested their framework only on animal data, it would be interesting to investigate its use on human data and eventually for BCI.

For BCI data, Guetschel et al.~\cite{GuePapTan21} developed a variation of the triplet loss that allows creating a hierarchical structure in the embedding space according to metadata associated with the recordings. In their framework, the hierarchy between the different meta-labels is defined by the researcher using expert knowledge. The authors demonstrated their framework by structuring the embedding space according to the subjects and motor imagery class labels, but this approach could theoretically be applied also to structure according to sessions, datasets or paradigms.

Studies exploiting metadata are at the border between supervised and unsupervised learning as they effectively use labels, but these metadata labels usually come "for free". 
On the other hand, contrastive losses can also be used to simply separate the classes as an alternative to a classification task~\cite{phunruangsakaoMultibranchConvolutionalNeural2022}.

Looking at the successful examples of deep metric learning techniques for face recognition, which works even for subjects outside the training set, we might wonder if this concept is not under-exploited for BCI data.
Indeed, we typically restrict ourselves to only a few classes in BCI, but the use of deep metric learning techniques could potentially deeply transform the BCI domain by, for example, allowing to learn from 24/7 multimodal recordings with both EEG and video for action recognition.


\section{Characterizing embeddings} \label{introspection}

In most studies, the introspection effort invested to characterize the learned representation is shallow and limited to simple score comparisons with a few baselines. 
This limited approach can often miss important details about the learned representations.
Fortunately, more elaborate techniques exist for introspecting DL-based embeddings, which can provide valuable insights into the learned representations. 
In this section, we will stress the importance of having common benchmarks with commonly agreed-on fine-tuning procedures to reliably compare a novel technique for obtaining an embedding with other existing techniques. 
We will also explain how the score on the pretext task can be used to better understand an embedding.
Finally, we will see projection methods for visualizing embeddings in lower dimensions, where it can be easier for humans to obtain insights. 


\subsection{Score on downstream task} \label{downstream-score}
While the performance an embedding enables for various tasks is a very high-level characteristic, it nevertheless is important. For testing how well an embedding will perform in transfer learning scenarios or to compare  \glspl{pretext-task}, specific benchmarks are required.
The MOABB library~\cite{aristimunhaMotherAllBCI2023} allows for rigorous benchmarking models on between-sessions and between-subject transfer scenarios with motor imagery, \gls{erp}, \acrshort{cvep} and \acrshort{ssvep} datasets~\cite{chevallierLargestEEGbasedBCI2024}.
However, these evaluations are not meant to allow fine-tuning the models on the target distribution.
This fine-tuning aspect was addressed in the 2021 BEETL competition~\cite{wei2021BEETLCompetition2022}.
In this competition, the participants had to solve a cross-datasets transfer task for motor imagery BCI data.
They received a few labelled examples also from the final test dataset, which makes this competition similar to a transfer-learning-with-fine-tuning task.
However, the participants of the BEETL competition were allowed to use the data as they wished.
Therefore, the challenge was simultaneously testing the initial training strategy and the eventual fine-tuning strategy of the participants.

\mysubsubsection{Intertwined evaluation}
In general, the performance score on a \gls{downstream-task} provides an intertwined evaluation of both the embedding method and the fine-tuning method.
Yet, the methods involved in both are independent and thus should be evaluated separately.
This separation is already common in the image domain~\cite{bardesVICRegVarianceInvarianceCovarianceRegularization2022,dengImageNetLargescaleHierarchical2009}.
To only test the initial training, i.e., compare different \glspl{pretext-task}, the typical approach is to always use the same fine-tuning strategy.
A first fine-tuning strategy is to continue the training of the whole network for a fixed number of epochs but on the \gls{downstream-task} and with examples from the target distribution~\cite{kostasBENDRUsingTransformers2021}. This strategy can be computationally heavy and may easily overfit depending on the amount of fine-tuning data.
 A second fine-tuning strategy also commonly used in representation learning is to train a linear classifier on top of the frozen representations~\cite{jingSelfSupervisedVisualFeature2021,bardesVICRegVarianceInvarianceCovarianceRegularization2022}.
This strategy produces reproducible results, and its simplicity favours methods able to extract representations which are easy to classify according to the \glspl{downstream-task}.
It thus may not obtain the best classification scores, but this is not crucial for benchmarking purposes.
Unfortunately, it is not always possible to use this so-called linear probing, depending on the \gls{ssl} strategy used~\cite{guetschelSJEPASeamlessCrossdataset2024}.

\mysubsubsection{Multiple downstream tasks}
Furthermore, a single \gls{downstream-task} is insufficient for the evaluation of models that claim a certain generalization characteristic, e.g., a EEG representation learned by unsupervised methods which shall be used for different BCI protocols as well as sleep staging, emotion recognition etc.
Instead, benchmarks containing sets of tasks are required, comparable to the benchmarks used in the image or language domain.
For \gls{nlp}, the GLUE benchmark~\cite{wangGLUEMultiTaskBenchmark2019} evaluates models on a range of natural language understanding tasks, 
while the SQuAD benchmark~\cite{rajpurkarSQuAD1000002016} evaluates models on question-answering tasks.
In computer vision, commonly used classification benchmarks are ImageNet~\cite{dengImageNetLargescaleHierarchical2009}, Places205~\cite{zhouLearningDeepFeatures2014}, VOC07~\cite{everinghamPascalVisualObject2010}, and iNat18~\cite{vanhornINaturalistSpeciesClassification2018};
and for object detection or segmentation, VOC07+12~\cite{everinghamPascalVisualObject2010}, and COCO~\cite{linMicrosoftCOCOCommon2014} are often used.
The development of these benchmarks has played a crucial role in advancing the field of \gls{nlp} and computer vision.

Similar benchmarks are needed for BCI to evaluate general-purpose embeddings.
In particular, such a benchmark would need to include all types of BCI tasks and should reflect a diversity of user groups, noise conditions, number and placements  of recording channels, recording qualities, number of calibration examples and contamination with artefacts.  
%
Aspects like the number of calibration examples, the number of channels, and the presence of noise or artefacts, can be simulated by applying corruptions or ablations to datasets. 
Such transformations are already used in studies for testing the robustness of models~\cite{nejedlyUtilizationTemporalAutoencoder2023,GuePapTan22,chienMAEEGMaskedAutoencoder2022,yangBIOTCrossdataBiosignal2023}. Unfortunately, the approaches are not consistent over publications, which makes comparison between studies difficult.
Normalizing these corruptions or ablations could be an option for establishing a standardized benchmark, allowing for more consistent and comparable evaluations across studies.

\mysubsubsection{Community adoption}
Finally, a good benchmark is one adopted by the community. 
If each article reports its results on a new benchmark, the authors should also provide baseline performances.
The high demands on computing resources in deep learning limits the number of baselines a new approach can be compared against. 
Additionally,  there is always a concern that authors applying a method as a baseline may not be using it to its fullest potential, whether intentionally or not.
For instance, a baseline method may be poorly optimized or implemented, leading to sub-optimal results.

In summary, the BCI community requires benchmarks that can evaluate general-purpose embeddings across a full range of tasks, have a deterministic fine-tuning procedure, and are widely adopted by the community for systematic model testing.

\subsection{Score on pretext task}

Obtaining a score evaluating the performance of a network on a \gls{pretext-task} is  straightforward, as each task comes with its own intrinsic metric.
It can simply be the value of the loss function or an accuracy score for tasks involving classification.
However, these scores are not ideal for comparing different \glspl{pretext-task} with each other because they are heterogeneous. 
Nonetheless, scores on \glspl{pretext-task} should not be disregarded, as they still provide important information which can complement \gls{downstream-task} scores. 
In particular, they can help with introspecting the embeddings learned by the model and their generalization abilities.
For example, in Banville et~al.~\cite{banvilleUncoveringStructureClinical2021}, plotting pretext and downstream task performances simultaneously allowed for a comparison of task difficulty and downstream benefits.
Kostas et~al.~\cite{kostasBENDRUsingTransformers2021} used the score on the pretext task to evaluate its difficulty with respect to its main hyperparameter. This relation between difficulty and hyperparameter allowed them to speculate on how the network was solving the task. Additionally, the low variability of the score on the pretext task across subjects, hardware, and tasks, allowed them to claim that the learned embeddings had good generalization abilities.
Overall, while pretext task scores should not be used to compare different pre-training methods together, they are useful for introspection during the development of a given pre-training method.

\subsection{Introspection by lower-dimensional visualizations of the embedding vectors}\label{projection}

\begin{figure}
\centering
\includegraphics[width=\linewidth]{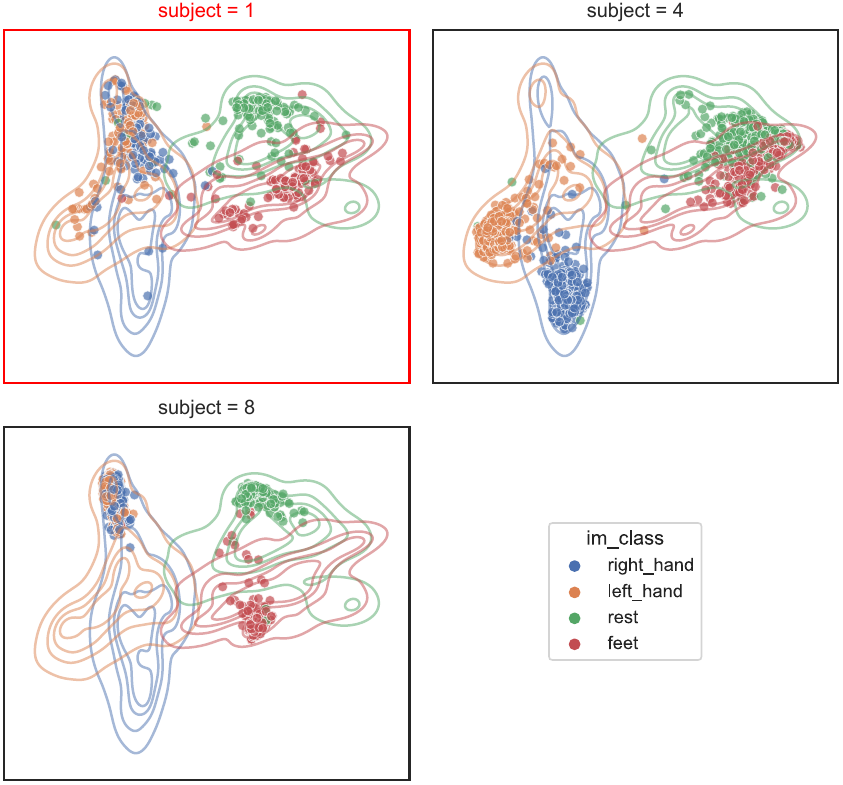}
\caption[Example of UMAP-projected embeddings]{
	Example of \gls{umap}-projected visualisation of embeddings. 
	\textit{Figure description.} In this figure, each point corresponds to one embedding vector. Its color denotes the class labels. 
	The sub-plots depict embedding vectors obtained for different subjects, but all vectors were generated by the same embedding function. The sub-plot of test subject 1 is marked by a red frame. The topographic isolines in the background indicate the four class distributions as derived from the complete data of all subjects. The plots of only three subjects are displayed here for space reasons.
	\textit{Comments.} This plot was used to realize that overall, the features learned were relevant for the targeted classification task, even for the test subject.
	Additionally, it indicated a hierarchy in the difficulty to separate the different pairs of classes. 
	The topographic isolines in the background served as a visual reference to compare sub-plots and allowed to observe distribution shifts between the embeddings of different subjects.
	\textit{Source:} Guetschel et~al.~2022~\cite{GuePapTan22}.	
}
\label{fig:umap}
\end{figure}

Embedding vectors typically have a few hundred dimensions.
Therefore, these vectors can not directly be visualised.
Thus it is common to first project embedding vectors into a two-dimensional space.
Then, all the examples of the dataset can be visualized simultaneously as a 2D scatter plot, each point representing a different example.
This type of plot allows to obtain insights about the distribution of the data in the original embedding space.
Additionally, it is common to colour the examples according to a label, typically the label corresponding to a \gls{downstream-task}~\cite{banvilleUncoveringStructureClinical2021,GuePapTan21,GuePapTan22,koMultiScaleNeuralNetwork2021,jeonMutualInformationDrivenSubjectInvariant2021}, 
but we sometimes encounter colourings corresponding to the age, gender, date, presence of a pathology~\cite{banvilleUncoveringStructureClinical2021}, continuous behavioural labels~\cite{schneiderLearnableLatentEmbeddings2022}, the subject id~\cite{GuePapTan21} or other meta-data.
A colouring can be applied to investigate how difficult it will probably be to separate the learned features according to the label.
The examples can also be coloured according to whether they belong to the train or the test set.
A comparison of the two distributions would allow inspecting potential non-stationarities between the two sets, which may impact the generalization abilities of the model. 
Such comparison is particularly beneficial for transfer learning scenarios.
An example of such a projected plot can be found in \autoref{fig:umap}.

Naturally, some information is lost in the projection. Thus the different projection methods are required to intrinsically make assumptions about the type of information that is important and should be preserved.
The following paragraphs describe the most commonly used methods for embedding vector projections.

\mysubsubsection{PCA}
A~well-established technique that linearly projects data into a new coordinate system is \gls{pca}~\cite{tippingProbabilisticPrincipalComponent1999}.
The coordinates of the new system are arranged in decreasing order of the variance, which the original data displays in each novel coordinate.
To reduce the dimensionality of the embeddings, one typically choses the first two dimensions of this new coordinate system.
Therefore, we see that \gls{pca} gives importance to the variance of the data: only the directions with the highest variance between the embedding vectors will be represented in the projection.

Because the projection is linear, the aspects of the original high-dimensional embedding space represented by it are faithful.
However, variance as a measure of importance for the dimensions may not be a relevant criterion to describe the data.
%
Nevertheless, \gls{pca} is the least computationally expensive one for visualising embeddings out of the three methods identified by this review~\cite{halkon.FindingStructureRandomness2011}.


\mysubsubsection{t-SNE}
The \gls{tsne}~\cite{maatenVisualizingDataUsing2008,hintonStochasticNeighborEmbedding2002}
is also a well-established, but non-linear method for dimensionality reduction.
It first models the embedding vectors as a graph where each node is one vector and where edges represent a pairwise similarity between vectors, i.e., a normalized version of their Euclidean distance.
Then, it builds a low-dimensional projection of each embedding vector along with a graph following similar principles as the original graph.
The projections are optimized such that the Kullback-Leibler divergences between the edge weights of their graph and those of the original graph are minimized, i.e., it enforces the graphs in both, the high- and the low-dimensional space to be similar. 

This procedure preserves local structures: embedding vectors which are close to each other will also be close to each other after the projection.
However, with the random initialisation of the projections originally proposed, \gls{tsne} does not allow to preserve the global structure~\cite{kobakInitializationCriticalPreserving2021}. Thus the distances between eventual clusters in the projected space should not be interpreted.

Finally, in current implementations, \gls{tsne} is significantly more computationally expensive than \gls{pca}.
It has a complexity of \(O(n^2)\) with \(n\) the number of embedding vectors and assuming that \(k\), the number of projected dimension, is small (\(k\le 3\))~\cite{maatenVisualizingDataUsing2008}.

\mysubsubsection{UMAP}
\Gls{umap}~\cite{mcinnesUMAPUniformManifold2020}
is the most recent of the three methods presented here.
It is very similar to \gls{tsne}, but it has been formulated using stronger mathematical principles to guide its design choices.
In particular, it is better than the original formulation of \gls{tsne} at preserving the global structure of the original embedding space as discussed by Oskolkov~\cite{oskolkovTSNEVsUMAP2020}
because of its initialisation and the choice of cross-entropy for the loss function.
Also, it is less computationally expensive than the original \gls{tsne} while still being significantly more expensive than \gls{pca}~\cite{mcinnesPerformanceComparisonDimension}. 
Its complexity is \(O(n^{1.14})\)~\cite{mcinnesUMAPUniformManifold2020}.
However, It was recently shown~\cite{kobakInitializationCriticalPreserving2021} that \gls{tsne} can preserve the global structure as well as \gls{umap} if its random initialisation is replaced with a \gls{pca} initialisation.
Also, recent optimized versions of \gls{tsne},
such as \gls{fitsne}~\cite{lindermanFastInterpolationbasedTSNE2019},
seem comparable to \gls{umap} in terms of speed when projecting data into 2D or 3D~\cite{kobakArtUsingTSNE2019}.
Still, \gls{umap} can effortlessly scale up with the projection dimension \(k\) whereas \gls{tsne}'s complexity grows exponentially with \(k\)~\cite{mcinnesUMAPUniformManifold2020}. Increasing \(k\) is not relevant for visualisation purposes but can be if we want to use these algorithms for dimensionality reduction, for example, before applying a clustering approach.

\subsection{Visual qualitative evaluation} 
A visual qualitative evaluation of embeddings can be achieved using conditioned generation techniques. These techniques involve a neural network, typically called \textit{generator}, whose objective is to generate artificial examples (i.e., EEG epochs) that are as similar as possible to real examples. In this context, the only information the generator receives about a target (real) example that is to be imitated, is an embedding vector that represents it. For this reason, we say that the generation process is \textit{conditioned} by an embedding vector.
Assuming the generator is properly trained, the similarity between the artificial and the real examples is limited by the information contained in the embedding vectors which act as a bottleneck. If a example is perfectly represented in its embedding vector (i.e., without information loss), then a well-trained generator will be able to reconstruct it exactly. If some information is lost by the embedding, however, then the generator can only "guess" the original input.
Visually comparing an artificial example and its target is a way to evaluate which information was lost and which was maintained in the corresponding embedding.
Views of these examples can also be visualised in the form of spatial patterns~\cite{liVariationalAutoencoderBased2019} of frequency spectra~\cite{yuAdaptiveEEGFeature2021}.

Autoencoders (c.f.~\autoref{autoencoder}) intrinsically train a generator and allow for this type of analysis without additional effort~\cite{liVariationalAutoencoderBased2019,yuAdaptiveEEGFeature2021}.
However, this is not the case for other embedding methods.
Border et al.~proposed to use diffusion models as generators and investigated this introspection method with images~\cite{bordesHighFidelityVisualization2022}.
However, this has not been explored yet, to our knowledge, with EEG embeddings and there have only been a few studies experimenting with diffusion models and EEG signals~\cite{tormaEEGWaveDenoisingDiffusion2023,aristimunhaSyntheticSleepEEG2023,kleinSynthesizingEEGSignals2024}.


\section{Discussion and Recommendations}
In the previous sections, we reported as factually as possible our findings on the methods that can or could be used to learn embeddings for BCI, along with reasons for why one would want to do so, and the methods available to introspect them. We will now synthesise those findings and extract the main outcomes and discuss them. We will close with recommendations for future research on deep representation learning for BCI and EEG.

Concerning the methods employed to learn embeddings by the articles we analysed, 
we found that a large majority were autoencoder-based, accounting for approximately half of the articles surveyed. This observation was not surprising as they are based on a quite straightforward principle which has been in existence for a long time and as many variants have been developed~\cite{bankAutoencoders2021} even if we could reported on three only in our study.
Furthermore, we found a significant number of \glspl{gan}, representing approximately a fourth of the surveyed articles. Again, this result was expected as \glspl{gan} used to be the state-of-the-art of generative models, before the arrival of diffusion models, and have been abundantly used for EEG generation. 
However, none of the surveyed articles employed \glspl{gan} for the purpose of EEG representation learning, leaving this area open for further investigation.
Finally, if we leave aside the studies that simply use the hidden layer of a classifier as embedding, the remaining methods were only observed sporadically. This shows how little the field of deep representation learning has been explored in BCI by now.
Regardless, we close this paragraph by reminding the reader that we did not conduct a systematic review so these percentages might not reflect the real distribution of the current research on deep representation learning for BCI; they should be taken with a grain of salt.

In addition to these findings, we make three primary observations: 
\textit{firstly}, in very few studies the authors were using representations in a transfer learning scenario. Yet, there is great potential: deep learning models shine in BCI transfer learning scenarios~\cite{wei2021BEETLCompetition2022}. 
Moreover, re-using a pre-trained representation can completely erase (in the case of linear probing~\cite{GuePapTan22}) or at least alleviate (in the case of fine-tuning~\cite{kostasBENDRUsingTransformers2021}) the cost of using deep learning techniques, which qualifies them for online BCI applications. 
\textit{Secondly}, when it comes to cross-dataset transfer learning, the authors of the reviewed articles all applied their own procedures for pre-training, fine-tuning and evaluating of models. This makes the comparison of methods difficult. To compare pre-training methods, the two other steps (fine-tuning and evaluating) should be fixed and standardised. To our knowledge there  currently is only one standard benchmark for cross-dataset transfer in BCI~\cite{wei2021BEETLCompetition2022} but it still leaves room for improvement as discussed in \autoref{downstream-score}.
\textit{Finally}, the authors often learned an embedding as a side product of the method they use rather than as an explicit objective. In most cases, they ignore the obtained embedding and continue with their primary task despite the large panel of introspection techniques available, as delineated in \autoref{introspection}.

In light of these observations, we proceed to sketch recommendations regarding the future of deep representation learning for EEG data.
The first recommendation is, when an embedding has been learned, to introspect this representation. Authors can choose from a number of existing techniques, as explained in \autoref{introspection}, that can all provide valuable insights about what has actually been learned. Moreover, this additional introspection step requires relatively little computational effort compared to the initial one for learning the embedding.

The second recommendation focuses on foundation models. We call \textit{foundation model} an architecture which has been pre-trained on large amounts of data. Such models can typically produce general representations of their input data. Foundation models serve as starting points or as building blocks for fine-tuning on \glspl{downstream-task}.
We believe that the development of EEG-specific foundation models would offer a great benefit for the BCI community.
In the computer vision domain, foundation models are typically trained using \gls{ssl} techniques~\cite{balestrieroCookbookSelfSupervisedLearning2023} so this seems to be a promising avenue for BCI, too.
Although the idea of developing foundation models for BCI or EEG has started to be discussed~\cite{cuiNeuroGPTDevelopingFoundation2023}, there currently is no such model that has been widely adopted by the BCI community.

Our third recommendation is about the eventual creation of novel EEG datasets for the training of foundation models. Experience from the language processing and the computer vision fields has shown that foundation models require  extremely large, but not necessarily labelled datasets to be trained~\cite{devlinBERTPretrainingDeep2019,alayracFlamingoVisualLanguage2022}. We assume this  will probably be the case also for EEG foundation models. In non-EEG domains, those large datasets were coming from very diverse sources, which would probably translate for EEG foundation models into many different EEG recording systems, subjects and recording conditions. The Temple University Hospital EEG data corpus~\cite{obeidTempleUniversityHospital2016} might be such a resource and has already been explored by Kostas et al.~\cite{kostasBENDRUsingTransformers2021} to train a \gls{ssl} model, but we still lack hindsight on whether this corpus is a good dataset for training foundation models.

Our final recommendation is about the eventual creation of novel benchmarks for evaluating \gls{ssl} methods and foundation models for BCI. To go beyond the BEETL benchmark~\cite{wei2021BEETLCompetition2022}, 
we first need to establish a set of fixed fine-tuning procedures to focus on a comparison between pre-trained models and not the combinations of pre-trained models and fine-tuning procedures. The choice of such fixed fine-tuning procedures should be relatively deterministic and realistic for BCI usage. Two simple fine-tuning procedures could for example be linear probing~\cite{GuePapTan22} and whole network fine-tuning~\cite{kostasBENDRUsingTransformers2021}.
Second, this new benchmark would need a large diversity of downstream tasks. As the main purpose of foundation models is to be re-used, they also need to be tested and compared on as many re-usage scenarios as possible. For BCI, this would translate into including datasets from as many BCI paradigms, recording scenarios and user groups as possible.  
We can even go a step further and mention that novel datasets could be recorded explicitly for this benchmark. As the goal of using a foundation model is to reduce the amount of calibration data needed, those novel datasets could probably contain less repetitions per condition but instead represent more and diverse conditions. 
Third, and finally, a benchmark is useful if it is adopted and used by the community. For this reason, the needs of the BCI community must be kept in mind. Also, it might be relevant to include such a benchmark in a tool already actively used by the community such as the MOABB library~\cite{aristimunhaMotherAllBCI2023}. An adoption would also be encouraged, if reviewers regularly require comparisons with baseline approaches.

The conceptualization and successful training of foundation models for EEG data processing, potentially facilitated by novel benchmarks and datasets, could revolutionize the field of BCI. 
In particular, it would reduce the amount of data needed to train BCI decoding models, implying reduced calibration times for novel sessions or subjects and a facilitation of rapidly explorating novel experimental paradigms and user tasks.
This efficiency could accelerate research cycles, potentially catalyzing the emergence of a new generation of BCI paradigms. For novel application fields of BCI such as invasive neurotechnological applications, where the small sample problems may be even more severe, this efficiency may even be decisive.
Furthermore, the generalized EEG representations we would obtain from future foundation models
could facilitate the alignment of brain signals with other data modalities, such as subject videos or medical records.
Such cross-modal embedding alignments would open the door for a broader scope of predictive tasks, extending beyond traditional imagery- and evoked potential-based paradigms to directly forecast attributes or states represented in other domains.

In summary, the advent of EEG foundation models could mark a paradigm shift, with implications ranging from streamlined model training and enhanced cross-modal applicability to the eradication of cumbersome calibration procedures. Therefore, a focus on establishing such models should be considered a priority within the EEG and BCI research communities and for funding decisions.

\section{Acknowledgements}

This work is in part supported by the Donders Center for Cognition (DCC) and is part of the project Dutch Brain Interface Initiative (DBI2) with project number 024.005.022 of the research program Gravitation which is (partly) financed by the Dutch Research Council (NWO).

\newpage
\glsaddall
\printnoidxglossary[type=\acronymtype]
\newpage
\printnoidxglossary

\newpage
\nocite{ayoobiSubjectindependentBraincomputerInterface2022}\nocite{banvilleUncoveringStructureClinical2021}\nocite{bashivanLearningRepresentationsEEG2016}\nocite{bruschMultiviewSelfsupervisedLearning2023}\nocite{chenCRETSCAENovelClassification2024}\nocite{chenDenoisingAutoencoderbasedFeature2021}\nocite{chienMAEEGMaskedAutoencoder2022}
\nocite{chuangConvolutionalDenoisingAutoencoder2022}\nocite{cuiNeuroGPTDevelopingFoundation2023}\nocite{ditthapronUniversalJointFeature2019}\nocite{doseEndtoendDeepLearning2018}\nocite{fahimiGenerativeAdversarialNetworksbased2021}\nocite{ferriStackedAutoencodersNew2021}\nocite{flintRepresentationFingerMovement2020}\nocite{foumaniEEG2RepEnhancingSelfsupervised2024}\nocite{gaoGenerativeAdversarialNetwork2022}\nocite{ghazikhaniStackedAutoencodersApproach2018}\nocite{GuePapTan21}\nocite{GuePapTan22}\nocite{guetschelSJEPASeamlessCrossdataset2024}\nocite{hartmannEEGGANGenerativeAdversarial2018}
\nocite{heSelfsupervisedLearningBased2022}\nocite{hwaidiClassificationMotorImagery2022}\nocite{jeonMutualInformationDrivenSubjectInvariant2021}\nocite{jiangApplicationTransformerAutoencoder2022}\nocite{jingweiDeepLearningEEG2015}\nocite{kanENHANCINGMULTICHANNELEEG2021}\nocite{kangMetaBCIPerspectivesRole2022}\nocite{koblerSPDDomainspecificBatch2022}\nocite{koMultiScaleNeuralNetwork2021}\nocite{kostasBENDRUsingTransformers2021}
\nocite{kumaraguruTrustAwareRouting2021}\nocite{leeMotorImageryClassification2022}\nocite{liLatentFactorDecoding2020}\nocite{liMultitaskCollaborativeNetwork2024}\nocite{linClassificationEpilepticEEG2016}\nocite{liuDistinguishableSpatialspectralFeature2021}\nocite{liuEEGBasedEmotionClassification2020}\nocite{liVariationalAutoencoderBased2019}\nocite{mammoneAutoEncoderFilterBank2023}
\nocite{mirzaeiEEGMotorImagery2021}\nocite{mohsenvandContrastiveRepresentationLearning2020}\nocite{nairImprovedApproachEEG2018}\nocite{nejedlyUtilizationTemporalAutoencoder2023}\nocite{ouImprovedSelfsupervisedLearning2022}\nocite{ozdenizciLearningInvariantRepresentations2020}\nocite{ozdenizciTransferLearningBrainComputer2019}
\nocite{parashivaNewChannelSelection2019}\nocite{parijaAutoencoderbasedImprovedDeep2022}\nocite{peiDecodingAsynchronousReaching2018}\nocite{petrutiuEnhancingClassificationEEG2020}\nocite{phadikarUnsupervisedFeatureExtraction2023}\nocite{phunruangsakaoMultibranchConvolutionalNeural2022}
\nocite{prabhakarSASDLRBATQSparse2022}\nocite{qiuDenoisingSparseAutoencoderbased2018}\nocite{ranHybridAutoencoderFramework2022}\nocite{rommelDeepInvariantNetworks2022}\nocite{royMIEEGGANGeneratingArtificial2020}\nocite{schneiderLearnableLatentEmbeddings2022}\nocite{songEEGConformerConvolutional2023}\nocite{stepheMotorImageryEEG2022}
\nocite{tanAutoencoderbasedTransferLearning2019}\nocite{tangMotorImageryEEG2019}\nocite{tomonagaExperimentsClassificationElectroencephalography2017}\nocite{wangNovelAlgorithmicStructure2023}\nocite{xieMotorImageryEEG2020}\nocite{xuRepresentationLearningMotor2021}\nocite{yanEEGSignalClassification2016}\nocite{yangAssessingCognitiveMental2019}\nocite{yangBIOTCrossdataBiosignal2023}
\nocite{yangNovelDeepLearning2020}\nocite{yangSelfsupervisedEEGRepresentation2023}\nocite{yaoEmotionClassificationBased2024}\nocite{yinPhysiologicalsignalbasedMentalWorkload2019}\nocite{yuAdaptiveEEGFeature2021}\nocite{zhangBrain2ObjectPrintingYour2020}
\nocite{zhangConvertingYourThoughts2018}\nocite{zhangMultitaskGenerativeAdversarial2020}\nocite{zhangRealizingApplicationEEG2021}\nocite{zhangSpectralTemporalFeature2019}\nocite{zhaoDeepCNNModel2022}

\entryneedsurldate{mcinnesPerformanceComparisonDimension}
\entryneedsurldate{oskolkovTSNEVsUMAP2020}
\printbibliography

@article{alayracFlamingoVisualLanguage2022,
  title = {Flamingo: A {{Visual Language Model}} for {{Few-Shot Learning}}},
  shorttitle = {Flamingo},
  author = {Alayrac, Jean-Baptiste and Donahue, Jeff and Luc, Pauline and Miech, Antoine and Barr, Iain and Hasson, Yana and Lenc, Karel and Mensch, Arthur and Millican, Katherine and Reynolds, Malcolm and Ring, Roman and Rutherford, Eliza and Cabi, Serkan and Han, Tengda and Gong, Zhitao and Samangooei, Sina and Monteiro, Marianne and Menick, Jacob L. and Borgeaud, Sebastian and Brock, Andy and Nematzadeh, Aida and Sharifzadeh, Sahand and Bi{\'n}kowski, Miko{\l}aj and Barreira, Ricardo and Vinyals, Oriol and Zisserman, Andrew and Simonyan, Kar{\'e}n},
  year = {2022},
  month = dec,
  journal = {Advances in Neural Information Processing Systems},
  volume = {35},
  pages = {23716--23736},
  urldate = {2023-08-23},
  langid = {english}
}

@inproceedings{arandjelovicNetVLADCNNArchitecture2016,
  title = {{{NetVLAD}}: {{CNN Architecture}} for {{Weakly Supervised Place Recognition}}},
  shorttitle = {{{NetVLAD}}},
  booktitle = {Proceedings of the {{IEEE Conference}} on {{Computer Vision}} and {{Pattern Recognition}}},
  author = {Arandjelovic, Relja and Gronat, Petr and Torii, Akihiko and Pajdla, Tomas and Sivic, Josef},
  year = {2016},
  pages = {5297--5307},
  urldate = {2023-03-02}
}

@misc{aristimunhaMotherAllBCI2023,
  title = {Mother of All {{BCI Benchmarks}}},
  author = {Aristimunha, Bruno and Carrara, Igor and Guetschel, Pierre and Sedlar, Sara and Rodrigues, Pedro and Sosulski, Jan and Narayanan, Divyesh and Bjareholt, Erik and Quentin, Barthelemy and Schirrmeister, Robin Tibor and Kalunga, Emmanuel and Darmet, Ludovic and Gregoire, Cattan and Abdul Hussain, Ali and Gatti, Ramiro and Goncharenko, Vladislav and Thielen, Jordy and Moreau, Thomas and Roy, Yannick and Jayaram, Vinay and Barachant, Alexandre and Chevallier, Sylvain},
  year = {2023},
  month = oct,
  doi = {10.5281/ZENODO.10034223},
  urldate = {2023-11-13},
  copyright = {Creative Commons Attribution 4.0 International},
  howpublished = {Zenodo}
}

@inproceedings{aristimunhaSyntheticSleepEEG2023,
  title = {Synthetic {{Sleep EEG Signal Generation}} Using {{Latent Diffusion Models}}},
  booktitle = {Deep {{Generative Models}} for {{Health Workshop NeurIPS}} 2023},
  author = {Aristimunha, Bruno and de Camargo, Raphael Yokoingawa and Chevallier, Sylvain and Lucena, Oeslle and Thomas, Adam G. and Cardoso, M. Jorge and Pinaya, Walter Hugo Lopez and Dafflon, Jessica},
  year = {2023},
  month = oct,
  urldate = {2024-04-09},
  langid = {english}
}

@inproceedings{ayoobiSubjectindependentBraincomputerInterface2022,
  type = {Journal Article},
  title = {A Subject-Independent Brain-Computer Interface Framework Based on Supervised Autoencoder.},
  booktitle = {Annual {{International Conference}} of the {{IEEE Engineering}} in {{Medicine}} and {{Biology Society}}},
  author = {Ayoobi, Navid and Sadeghian, Elnaz Banan},
  year = {2022},
  pages = {218--221},
  doi = {10.1109/embc48229.2022.9871590}
}

@inproceedings{baevskiWav2vecFrameworkSelfSupervised2020,
  title = {Wav2vec 2.0: {{A Framework}} for {{Self-Supervised Learning}} of {{Speech Representations}}},
  shorttitle = {Wav2vec 2.0},
  booktitle = {Advances in {{Neural Information Processing Systems}}},
  author = {Baevski, Alexei and Zhou, Yuhao and Mohamed, Abdelrahman and Auli, Michael},
  year = {2020},
  volume = {33},
  pages = {12449--12460},
  publisher = {Curran Associates, Inc.},
  urldate = {2023-02-03}
}

@misc{balestrieroCookbookSelfSupervisedLearning2023,
  title = {A {{Cookbook}} of {{Self-Supervised Learning}}},
  author = {Balestriero, Randall and Ibrahim, Mark and Sobal, Vlad and Morcos, Ari and Shekhar, Shashank and Goldstein, Tom and Bordes, Florian and Bardes, Adrien and Mialon, Gregoire and Tian, Yuandong and Schwarzschild, Avi and Wilson, Andrew Gordon and Geiping, Jonas and Garrido, Quentin and Fernandez, Pierre and Bar, Amir and Pirsiavash, Hamed and LeCun, Yann and Goldblum, Micah},
  year = {2023},
  month = apr,
  number = {arXiv:2304.12210},
  eprint = {2304.12210},
  primaryclass = {cs},
  publisher = {arXiv},
  doi = {10.48550/arXiv.2304.12210},
  urldate = {2023-04-28},
  archiveprefix = {arxiv}
}

@misc{bankAutoencoders2021,
  title = {Autoencoders},
  author = {Bank, Dor and Koenigstein, Noam and Giryes, Raja},
  year = {2021},
  month = apr,
  number = {arXiv:2003.05991},
  eprint = {2003.05991},
  primaryclass = {cs, stat},
  publisher = {arXiv},
  urldate = {2023-07-27},
  archiveprefix = {arxiv}
}

@article{banvilleUncoveringStructureClinical2021,
  title = {Uncovering the Structure of Clinical {{EEG}} Signals with Self-Supervised Learning},
  author = {Banville, Hubert and Chehab, Omar and Hyv{\"a}rinen, Aapo and Engemann, Denis-Alexander and Gramfort, Alexandre},
  year = {2021},
  month = aug,
  journal = {Journal of Neural Engineering},
  volume = {18},
  number = {4},
  pages = {046020},
  issn = {1741-2560, 1741-2552},
  doi = {10.1088/1741-2552/abca18},
  urldate = {2022-05-04},
  langid = {english}
}

@misc{bardesVICRegVarianceInvarianceCovarianceRegularization2022,
  title = {{{VICReg}}: {{Variance-Invariance-Covariance Regularization}} for {{Self-Supervised Learning}}},
  shorttitle = {{{VICReg}}},
  author = {Bardes, Adrien and Ponce, Jean and LeCun, Yann},
  year = {2022},
  month = jan,
  number = {arXiv:2105.04906},
  eprint = {2105.04906},
  primaryclass = {cs},
  publisher = {arXiv},
  doi = {10.48550/arXiv.2105.04906},
  urldate = {2022-12-16},
  archiveprefix = {arxiv}
}

@misc{bashivanLearningRepresentationsEEG2016,
  title = {Learning {{Representations}} from {{EEG}} with {{Deep Recurrent-Convolutional Neural Networks}}},
  author = {Bashivan, Pouya and Rish, Irina and Yeasin, Mohammed and Codella, Noel},
  year = {2016},
  month = feb,
  number = {arXiv:1511.06448},
  eprint = {1511.06448},
  primaryclass = {cs},
  publisher = {arXiv},
  doi = {10.48550/arXiv.1511.06448},
  urldate = {2022-12-05},
  archiveprefix = {arxiv}
}

@misc{bordesHighFidelityVisualization2022,
  title = {High {{Fidelity Visualization}} of {{What Your Self-Supervised Representation Knows About}}},
  author = {Bordes, Florian and Balestriero, Randall and Vincent, Pascal},
  year = {2022},
  month = aug,
  number = {arXiv:2112.09164},
  eprint = {2112.09164},
  primaryclass = {cs},
  publisher = {arXiv},
  doi = {10.48550/arXiv.2112.09164},
  urldate = {2023-07-03},
  archiveprefix = {arxiv}
}

@misc{bruschMultiviewSelfsupervisedLearning2023,
  title = {Multi-View Self-Supervised Learning for Multivariate Variable-Channel Time Series},
  author = {Br{\"u}sch, Thea and Schmidt, Mikkel N. and Alstr{\o}m, Tommy S.},
  year = {2023},
  month = jul,
  number = {arXiv:2307.09614},
  eprint = {2307.09614},
  primaryclass = {cs, eess, stat},
  publisher = {arXiv},
  doi = {10.48550/arXiv.2307.09614},
  urldate = {2023-10-03},
  archiveprefix = {arxiv}
}

@article{castano-candamilPosthocLabelingArbitrary2019,
  title = {Post-Hoc {{Labeling}} of {{Arbitrary M}}/{{EEG Recordings}} for {{Data-Efficient Evaluation}} of {{Neural Decoding Methods}}},
  author = {{Casta{\~n}o-Candamil}, Sebasti{\'a}n and Meinel, Andreas and Tangermann, Michael},
  year = {2019},
  journal = {Frontiers in Neuroinformatics},
  volume = {13},
  issn = {1662-5196},
  doi = {10.3389/fninf.2019.00055},
  urldate = {2023-07-28}
}

@article{chenCRETSCAENovelClassification2024,
  type = {Journal Article},
  title = {{{CRE-TSCAE}}: {{A}} Novel Classification Model Based on Stacked Convolutional Autoencoder for Dual-Target {{RSVP-BCI}} Tasks},
  author = {Chen, Hongying and Wang, Dan and Xu, Meng and Chen, Yuanfang},
  year = {2024},
  journal = {IEEE transactions on bio-medical engineering},
  issn = {1558-2531},
  doi = {10.1109/TBME.2024.3361716}
}

@article{chenDenoisingAutoencoderbasedFeature2021,
  type = {Journal Article},
  title = {Denoising Autoencoder-Based Feature Extraction to Robust {{SSVEP-Based BCIs}}.},
  author = {Chen, Yeou-Jiunn and Chen, Pei-Chung and Chen, Shih-Chung and Wu, Chung-Min},
  year = {2021},
  journal = {Sensors (Basel, Switzerland)},
  volume = {21},
  number = {15},
  issn = {1424-8220},
  doi = {10.3390/s21155019}
}

@misc{chenSimpleFrameworkContrastive2020,
  title = {A {{Simple Framework}} for {{Contrastive Learning}} of {{Visual Representations}}},
  author = {Chen, Ting and Kornblith, Simon and Norouzi, Mohammad and Hinton, Geoffrey},
  year = {2020},
  month = jun,
  number = {arXiv:2002.05709},
  eprint = {2002.05709},
  primaryclass = {cs, stat},
  publisher = {arXiv},
  doi = {10.48550/arXiv.2002.05709},
  urldate = {2023-02-27},
  archiveprefix = {arxiv}
}

@misc{chevallierLargestEEGbasedBCI2024,
  title = {The Largest {{EEG-based BCI}} Reproducibility Study for Open Science: The {{MOABB}} Benchmark},
  shorttitle = {The Largest {{EEG-based BCI}} Reproducibility Study for Open Science},
  author = {Chevallier, Sylvain and Carrara, Igor and Aristimunha, Bruno and Guetschel, Pierre and Sedlar, Sara and Lopes, Bruna and Velut, Sebastien and Khazem, Salim and Moreau, Thomas},
  year = {2024},
  month = apr,
  number = {arXiv:2404.15319},
  eprint = {2404.15319},
  primaryclass = {cs, eess, q-bio},
  publisher = {arXiv},
  doi = {10.48550/arXiv.2404.15319},
  urldate = {2024-04-29},
  archiveprefix = {arxiv}
}

@misc{chienMAEEGMaskedAutoencoder2022,
  title = {{{MAEEG}}: {{Masked Auto-encoder}} for {{EEG Representation Learning}}},
  shorttitle = {{{MAEEG}}},
  author = {Chien, Hsiang-Yun Sherry and Goh, Hanlin and Sandino, Christopher M. and Cheng, Joseph Y.},
  year = {2022},
  month = oct,
  number = {arXiv:2211.02625},
  eprint = {2211.02625},
  primaryclass = {cs, eess},
  publisher = {arXiv},
  doi = {10.48550/arXiv.2211.02625},
  urldate = {2023-10-03},
  archiveprefix = {arxiv}
}

@article{chuangConvolutionalDenoisingAutoencoder2022,
  type = {Article},
  title = {Convolutional Denoising Autoencoder Based {{SSVEP}} Signal Enhancement to {{SSVEP-based BCIs}}},
  author = {Chuang, Chia-Chun and Lee, Chien-Ching and Yeng, Chia-Hong and So, Edmund-Cheung and Lin, Bor-Shyh and Chen, Yeou-Jiunn},
  year = {2022},
  journal = {Microsystem Technologies-Micro-And Nanosystems-Information Storage And Processing Systems},
  volume = {28},
  number = {1},
  pages = {237--244},
  issn = {0946-7076},
  doi = {10.1007/s00542-019-04654-2}
}

@misc{cuiNeuroGPTDevelopingFoundation2023,
  title = {Neuro-{{GPT}}: {{Developing A Foundation Model}} for {{EEG}}},
  shorttitle = {Neuro-{{GPT}}},
  author = {Cui, Wenhui and Jeong, Woojae and Th{\"o}lke, Philipp and Medani, Takfarinas and Jerbi, Karim and Joshi, Anand A. and Leahy, Richard M.},
  year = {2023},
  month = nov,
  number = {arXiv:2311.03764},
  eprint = {2311.03764},
  primaryclass = {cs, eess},
  publisher = {arXiv},
  doi = {10.48550/arXiv.2311.03764},
  urldate = {2023-11-15},
  archiveprefix = {arxiv}
}

@inproceedings{dengImageNetLargescaleHierarchical2009,
  title = {{{ImageNet}}: {{A}} Large-Scale Hierarchical Image Database},
  shorttitle = {{{ImageNet}}},
  booktitle = {{{IEEE Conference}} on {{Computer Vision}} and {{Pattern Recognition}}},
  author = {Deng, Jia and Dong, Wei and Socher, Richard and Li, Li-Jia and Li, Kai and {Fei-Fei}, Li},
  year = {2009},
  month = jun,
  pages = {248--255},
  issn = {1063-6919},
  doi = {10.1109/CVPR.2009.5206848}
}

@misc{devlinBERTPretrainingDeep2019,
  title = {{{BERT}}: {{Pre-training}} of {{Deep Bidirectional Transformers}} for {{Language Understanding}}},
  shorttitle = {{{BERT}}},
  author = {Devlin, Jacob and Chang, Ming-Wei and Lee, Kenton and Toutanova, Kristina},
  year = {2019},
  month = may,
  number = {arXiv:1810.04805},
  eprint = {1810.04805},
  primaryclass = {cs},
  publisher = {arXiv},
  doi = {10.48550/arXiv.1810.04805},
  urldate = {2023-02-27},
  archiveprefix = {arxiv}
}

@article{ditthapronUniversalJointFeature2019,
  title = {Universal {{Joint Feature Extraction}} for {{P300 EEG Classification Using Multi-Task Autoencoder}}},
  author = {Ditthapron, Apiwat and Banluesombatkul, Nannapas and Ketrat, Sombat and Chuangsuwanich, Ekapol and Wilaiprasitporn, Theerawit},
  year = {2019},
  journal = {IEEE Access},
  volume = {7},
  pages = {68415--68428},
  issn = {2169-3536},
  doi = {10.1109/ACCESS.2019.2919143}
}

@article{doseEndtoendDeepLearning2018,
  title = {An End-to-End Deep Learning Approach to {{MI-EEG}} Signal Classification for {{BCIs}}},
  author = {Dose, Hauke and M{\o}ller, Jakob S. and Iversen, Helle K. and Puthusserypady, Sadasivan},
  year = {2018},
  month = dec,
  journal = {Expert Systems with Applications},
  volume = {114},
  pages = {532--542},
  issn = {0957-4174},
  doi = {10.1016/j.eswa.2018.08.031},
  urldate = {2022-10-04},
  langid = {english}
}

@article{everinghamPascalVisualObject2010,
  title = {The {{Pascal Visual Object Classes}} ({{VOC}}) {{Challenge}}},
  author = {Everingham, Mark and Van Gool, Luc and Williams, Christopher K. I. and Winn, John and Zisserman, Andrew},
  year = {2010},
  month = jun,
  journal = {International Journal of Computer Vision},
  volume = {88},
  number = {2},
  pages = {303--338},
  issn = {1573-1405},
  doi = {10.1007/s11263-009-0275-4},
  urldate = {2023-04-20},
  langid = {english}
}

@article{fahimiGenerativeAdversarialNetworksbased2021,
  type = {Journal Article},
  title = {Generative Adversarial Networks-Based Data Augmentation for Brain-Computer Interface.},
  author = {Fahimi, Fatemeh and Dosen, Strahinja and Ang, Kai Keng and {Mrachacz-Kersting}, Natalie and Guan, Cuntai},
  year = {2021},
  journal = {IEEE transactions on neural networks and learning systems},
  volume = {32},
  number = {9},
  pages = {4039--4051},
  issn = {2162-2388},
  doi = {10.1109/tnnls.2020.3016666}
}

@article{ferriStackedAutoencodersNew2021,
  type = {Journal Article},
  title = {Stacked Autoencoders as New Models for an Accurate {{Alzheimer}}'s Disease Classification Support Using Resting-State {{EEG}} and {{MRI}} Measurements.},
  author = {Ferri, Raffaele and Babiloni, Claudio and Karami, Vania and Triggiani, Antonio Ivano and Carducci, Filippo and Noce, Giuseppe and Lizio, Roberta and Pascarelli, Maria T and Soricelli, Andrea and Amenta, Francesco and Bozzao, Alessandro and Romano, Andrea and Giubilei, Franco and Percio, Claudio Del and Stocchi, Fabrizio and Frisoni, Giovanni B and Nobili, Flavio and Patan{\`e}, Luca and Arena, Paolo},
  year = {2021},
  journal = {Clinical neurophysiology : official journal of the International Federation of Clinical Neurophysiology},
  volume = {132},
  number = {1},
  pages = {232--245},
  issn = {1872-8952},
  doi = {10.1016/j.clinph.2020.09.015}
}

@article{flintRepresentationFingerMovement2020,
  title = {The {{Representation}} of {{Finger Movement}} and {{Force}} in {{Human Motor}} and {{Premotor Cortices}}},
  author = {Flint, Robert D. and Tate, Matthew C. and Li, Kejun and Templer, Jessica W. and Rosenow, Joshua M. and Pandarinath, Chethan and Slutzky, Marc W.},
  year = {2020},
  month = aug,
  journal = {eNeuro},
  volume = {7},
  number = {4},
  issn = {2373-2822},
  doi = {10.1523/ENEURO.0063-20.2020},
  urldate = {2022-10-03},
  pmcid = {PMC7438059},
  pmid = {32769159}
}

@misc{foumaniEEG2RepEnhancingSelfsupervised2024,
  title = {{{EEG2Rep}}: {{Enhancing Self-supervised EEG Representation Through Informative Masked Inputs}}},
  shorttitle = {{{EEG2Rep}}},
  author = {Foumani, Navid Mohammadi and Mackellar, Geoffrey and Ghane, Soheila and Irtza, Saad and Nguyen, Nam and Salehi, Mahsa},
  year = {2024},
  month = feb,
  number = {arXiv:2402.17772},
  eprint = {2402.17772},
  primaryclass = {cs, eess},
  publisher = {arXiv},
  doi = {10.48550/arXiv.2402.17772},
  urldate = {2024-03-02},
  archiveprefix = {arxiv}
}

@inproceedings{ganinUnsupervisedDomainAdaptation2015,
  title = {Unsupervised {{Domain Adaptation}} by {{Backpropagation}}},
  booktitle = {Proceedings of the 32nd {{International Conference}} on {{Machine Learning}}},
  author = {Ganin, Yaroslav and Lempitsky, Victor},
  year = {2015},
  month = jun,
  pages = {1180--1189},
  publisher = {PMLR},
  issn = {1938-7228},
  urldate = {2022-12-20},
  langid = {english}
}

@article{gaoGenerativeAdversarialNetwork2022,
  type = {Article},
  title = {Generative Adversarial Network and Convolutional Neural Network-Based {{EEG}} Imbalanced Classification Model for Seizure Detection},
  author = {Gao, Bin and Zhou, Jiazheng and Yang, Yuying and Chi, Jinxin and Yuan, Qi},
  year = {2022},
  journal = {Biocybernetics And Biomedical Engineering},
  volume = {42},
  number = {1},
  pages = {1--15},
  issn = {0208-5216},
  doi = {10.1016/j.bbe.2021.11.002}
}

@inproceedings{ghazikhaniStackedAutoencodersApproach2018,
  type = {Proceedings Paper},
  title = {A Stacked Autoencoders Approach for a {{P300}} Speller {{BCI}}},
  shorttitle = {{{ICCKE}}},
  booktitle = {8th {{International Conference On Computer And Knowledge Engineering}}},
  author = {Ghazikhani, Hamed and Rouhani, Modjtaba},
  year = {2018},
  pages = {1--6},
  doi = {10.1109/iccke.2018.8566534}
}

@article{goodfellowGenerativeAdversarialNetworks2020,
  title = {Generative Adversarial Networks},
  author = {Goodfellow, Ian and {Pouget-Abadie}, Jean and Mirza, Mehdi and Xu, Bing and {Warde-Farley}, David and Ozair, Sherjil and Courville, Aaron and Bengio, Yoshua},
  year = {2020},
  month = oct,
  journal = {Communications of the ACM},
  volume = {63},
  number = {11},
  pages = {139--144},
  issn = {0001-0782},
  doi = {10.1145/3422622},
  urldate = {2022-12-20}
}

@misc{GuePapTan21,
  title = {An Embedding for {{EEG}} Signals Learned Using a Triplet Loss},
  author = {Guetschel, Pierre and Papadopoulo, Th{\'e}odore and Tangermann, Michael},
  year = {2023},
  month = mar,
  number = {arXiv:2304.06495},
  eprint = {2304.06495},
  primaryclass = {cs, eess},
  publisher = {arXiv},
  doi = {10.48550/ARXIV.2304.06495},
  archiveprefix = {arxiv},
  copyright = {All rights reserved},
  langid = {english}
}

@inproceedings{GuePapTan22,
  title = {Embedding Neurophysiological Signals},
  booktitle = {International {{Conference}} on {{Metrology}} for {{eXtended Reality}}, {{Artificial Intelligence}}, and {{Neural Engineering}} ({{MetroXRAINE}})},
  author = {Guetschel, Pierre and Papadopoulo, Thoedore and Tangermann, Michael},
  year = {2022},
  month = oct,
  pages = {169--174},
  publisher = {IEEE},
  address = {Rome},
  doi = {10.1109/metroxraine54828.2022.9967496},
  copyright = {All rights reserved},
  langid = {english}
}

@misc{guetschelSJEPASeamlessCrossdataset2024,
  title = {S-{{JEPA}}: Towards Seamless Cross-Dataset Transfer through Dynamic Spatial Attention},
  shorttitle = {S-{{JEPA}}},
  author = {Guetschel, Pierre and Moreau, Thomas and Tangermann, Michael},
  year = {2024},
  month = mar,
  number = {arXiv:2403.11772},
  eprint = {2403.11772},
  primaryclass = {cs},
  publisher = {arXiv},
  doi = {10.48550/arXiv.2403.11772},
  urldate = {2024-04-04},
  archiveprefix = {arxiv}
}

@inproceedings{hadsellDimensionalityReductionLearning2006,
  title = {Dimensionality {{Reduction}} by {{Learning}} an {{Invariant Mapping}}},
  booktitle = {{{IEEE Computer Society Conference}} on {{Computer Vision}} and {{Pattern Recognition}} ({{CVPR}}'06)},
  author = {Hadsell, R. and Chopra, S. and LeCun, Y.},
  year = {2006},
  month = jun,
  volume = {2},
  pages = {1735--1742},
  issn = {1063-6919},
  doi = {10.1109/CVPR.2006.100}
}

@article{halkon.FindingStructureRandomness2011,
  title = {Finding {{Structure}} with {{Randomness}}: {{Stochastic Algorithms}} for {{Constructing Approximate}} Matrix {{Decompositions}}},
  shorttitle = {Finding {{Structure}} with {{Randomness}}},
  author = {Halko, N. and Martinsson, P. G. and Tropp, J. A.},
  year = {2011},
  month = oct,
  journal = {California Institute of Technology},
  publisher = {California Institute of Technology},
  doi = {10.7907/PK8V-V047},
  urldate = {2023-03-28},
  collaborator = {Applied \& Computational Mathematics},
  copyright = {No commercial reproduction, distribution, display or performance rights in this work are provided.},
  langid = {english}
}

@misc{hartmannEEGGANGenerativeAdversarial2018,
  title = {{{EEG-GAN}}: {{Generative}} Adversarial Networks for Electroencephalograhic ({{EEG}}) Brain Signals},
  shorttitle = {{{EEG-GAN}}},
  author = {Hartmann, Kay Gregor and Schirrmeister, Robin Tibor and Ball, Tonio},
  year = {2018},
  month = jun,
  number = {arXiv:1806.01875},
  eprint = {1806.01875},
  primaryclass = {cs, eess, q-bio, stat},
  publisher = {arXiv},
  doi = {10.48550/arXiv.1806.01875},
  urldate = {2022-12-05},
  archiveprefix = {arxiv}
}

@misc{harzing-PublishPerish2007,
  title = {Publish or {{Perish}}},
  author = {{Harzing -}, Anne-Wil},
  year = {2007},
  urldate = {2023-07-10}
}

@article{heSelfsupervisedLearningBased2022,
  type = {Journal Article},
  title = {A Self-Supervised Learning Based Channel Attention {{MLP-Mixer}} Network for Motor Imagery Decoding.},
  author = {He, Yanbin and Lu, Zhiyang and Wang, Jun and Ying, Shihui and Shi, Jun},
  year = {2022},
  journal = {IEEE transactions on neural systems and rehabilitation engineering},
  volume = {30},
  pages = {2406--2417},
  issn = {1558-0210},
  doi = {10.1109/tnsre.2022.3199363}
}

@misc{higginsEarlyVisualConcept2016,
  title = {Early {{Visual Concept Learning}} with {{Unsupervised Deep Learning}}},
  author = {Higgins, Irina and Matthey, Loic and Glorot, Xavier and Pal, Arka and Uria, Benigno and Blundell, Charles and Mohamed, Shakir and Lerchner, Alexander},
  year = {2016},
  month = sep,
  number = {arXiv:1606.05579},
  eprint = {1606.05579},
  primaryclass = {cs, q-bio, stat},
  publisher = {arXiv},
  doi = {10.48550/arXiv.1606.05579},
  urldate = {2023-03-06},
  archiveprefix = {arxiv}
}

@inproceedings{hintonStochasticNeighborEmbedding2002,
  title = {Stochastic {{Neighbor Embedding}}},
  booktitle = {Advances in {{Neural Information Processing Systems}}},
  author = {Hinton, Geoffrey E and Roweis, Sam},
  year = {2002},
  volume = {15},
  publisher = {MIT Press},
  urldate = {2023-03-28}
}

@article{hwaidiClassificationMotorImagery2022,
  type = {Article},
  title = {Classification of Motor Imagery {{EEG}} Signals Based on Deep Autoencoder and Convolutional Neural Network Approach},
  author = {Hwaidi, Jamal F. and Chen, Thomas M.},
  year = {2022},
  journal = {IEEE access : practical innovations, open solutions},
  volume = {10},
  pages = {48071--48081},
  issn = {2169-3536},
  doi = {10.1109/access.2022.3171906}
}

@article{jayaramTransferLearningBraincomputer2016,
  title = {Transfer Learning in Brain-Computer Interfaces},
  author = {Jayaram, Vinay and Alamgir, Morteza and Altun, Yasemin and Scholkopf, Bernhard and {Grosse-Wentrup}, Moritz},
  year = {2016},
  journal = {IEEE Computational Intelligence Magazine},
  volume = {11},
  number = {1},
  pages = {20--31},
  publisher = {IEEE},
  issn = {1556-603X},
  doi = {10.1109/MCI.2015.2501545}
}

@article{jeonMutualInformationDrivenSubjectInvariant2021,
  title = {Mutual {{Information-Driven Subject-Invariant}} and {{Class-Relevant Deep Representation Learning}} in {{BCI}}},
  author = {Jeon, Eunjin and Ko, Wonjun and Yoon, Jee Seok and Suk, Heung-Il},
  year = {2021},
  journal = {IEEE Transactions on Neural Networks and Learning Systems},
  pages = {1--11},
  issn = {2162-237X, 2162-2388},
  doi = {10.1109/TNNLS.2021.3100583},
  urldate = {2022-09-19}
}

@inproceedings{jiangApplicationTransformerAutoencoder2022,
  type = {Proceedings Paper},
  title = {Application of Transformer with Auto-Encoder in Motor Imagery {{EEG}} Signals},
  booktitle = {14th {{International Conference}} on {{Wireless Communication}} and {{Signal Processing}}},
  author = {Jiang, Rui and Sun, Liuting and Wang, Xiaoming and Xu, Youyun},
  year = {2022},
  pages = {631--637},
  issn = {2325-3746},
  doi = {10.1109/WCSP55476.2022.10039415},
  affiliation = {Jiang, R (Corresponding Author), Nanjing Univ Posts \& Telecommun, Coll Telecommun \& Informat Engn, Nanjing, Jiangsu, Peoples R China. Jiang, Rui; Sun, Liuting; Wang, Xiaoming; Xu, Youyun, Nanjing Univ Posts \& Telecommun, Coll Telecommun \& Informat Engn, Nanjing, Jiangsu, Peoples R China.},
  author-email = {j\_ray@njupt.edu.cn},
  book-group-author = {IEEE},
  da = {2024-04-03},
  isbn = {978-1-66545-085-0},
  times-cited = {1},
  unique-id = {WOS:000972901000116}
}

@article{jingSelfSupervisedVisualFeature2021,
  title = {Self-{{Supervised Visual Feature Learning With Deep Neural Networks}}: {{A Survey}}},
  shorttitle = {Self-{{Supervised Visual Feature Learning With Deep Neural Networks}}},
  author = {Jing, Longlong and Tian, Yingli},
  year = {2021},
  month = nov,
  journal = {IEEE Transactions on Pattern Analysis and Machine Intelligence},
  volume = {43},
  number = {11},
  pages = {4037--4058},
  issn = {1939-3539},
  doi = {10.1109/TPAMI.2020.2992393}
}

@inproceedings{jingweiDeepLearningEEG2015,
  title = {Deep Learning {{EEG}} Response Representation for Brain Computer Interface},
  booktitle = {2015 34th {{Chinese Control Conference}} ({{CCC}})},
  author = {Jingwei, Liu and Yin, Cheng and Weidong, Zhang},
  year = {2015},
  month = jul,
  pages = {3518--3523},
  publisher = {IEEE},
  address = {Hangzhou, China},
  doi = {10.1109/ChiCC.2015.7260182},
  urldate = {2022-09-19},
  isbn = {978-988-15638-9-7}
}

@article{kanEnhancingMultichannelEEG2021,
  type = {Proceedings Paper},
  title = {Enhancing Multi-Channel {{EEG}} Classification with {{Gramian}} Temporal Generative Adversarial Networks},
  author = {Kan, Chi Nok Enoch and Povinelli, Richard J. and Ye, Dong Hye},
  year = {2021},
  journal = {IEEE International Conference On Acoustics, Speech And Signal Processing},
  pages = {1260--1264},
  doi = {10.1109/icassp39728.2021.9414078}
}

@inproceedings{kangMetaBCIPerspectivesRole2022,
  type = {Proceedings Paper},
  title = {Meta-{{BCI}}: {{Perspectives}} on a Role of Self-Supervised Learning in Meta Brain Computer Interface},
  booktitle = {10th {{International Winter Conference On Brain-Computer Interface}} ({{Bci2022}})},
  author = {Kang, Young Ho and Kim, Dongjae and Lee, Sang Wan},
  year = {2022},
  doi = {10.1109/bci53720.2022.9734995}
}

@misc{kirosUnifyingVisualSemanticEmbeddings2014,
  title = {Unifying {{Visual-Semantic Embeddings}} with {{Multimodal Neural Language Models}}},
  author = {Kiros, Ryan and Salakhutdinov, Ruslan and Zemel, Richard S.},
  year = {2014},
  publisher = {arXiv},
  doi = {10.48550/ARXIV.1411.2539},
  urldate = {2023-03-21},
  copyright = {arXiv.org perpetual, non-exclusive license}
}

@misc{kleinSynthesizingEEGSignals2024,
  title = {Synthesizing {{EEG Signals}} from {{Event-Related Potential Paradigms}} with {{Conditional Diffusion Models}}},
  author = {Klein, Guido and Guetschel, Pierre and Silvestri, Gianluigi and Tangermann, Michael},
  year = {2024},
  month = mar,
  number = {arXiv:2403.18486},
  eprint = {2403.18486},
  primaryclass = {cs, eess},
  publisher = {arXiv},
  doi = {10.48550/arXiv.2403.18486},
  urldate = {2024-04-09},
  archiveprefix = {arxiv}
}

@article{kobakArtUsingTSNE2019,
  title = {The Art of Using T-{{SNE}} for Single-Cell Transcriptomics},
  author = {Kobak, Dmitry and Berens, Philipp},
  year = {2019},
  month = nov,
  journal = {Nature Communications},
  volume = {10},
  number = {1},
  pages = {5416},
  publisher = {Nature Publishing Group},
  issn = {2041-1723},
  doi = {10.1038/s41467-019-13056-x},
  urldate = {2023-03-30},
  copyright = {2019 The Author(s)},
  langid = {english}
}

@article{kobakInitializationCriticalPreserving2021,
  title = {Initialization Is Critical for Preserving Global Data Structure in Both T-{{SNE}} and {{UMAP}}},
  author = {Kobak, Dmitry and Linderman, George C.},
  year = {2021},
  month = feb,
  journal = {Nature Biotechnology},
  volume = {39},
  number = {2},
  pages = {156--157},
  publisher = {Nature Publishing Group},
  issn = {1546-1696},
  doi = {10.1038/s41587-020-00809-z},
  urldate = {2023-03-30},
  copyright = {2021 The Author(s), under exclusive licence to Springer Nature America, Inc.},
  langid = {english}
}

@misc{koblerSPDDomainspecificBatch2022,
  title = {{{SPD}} Domain-Specific Batch Normalization to Crack Interpretable Unsupervised Domain Adaptation in {{EEG}}},
  author = {Kobler, Reinmar J. and Hirayama, Jun-ichiro and Zhao, Qibin and Kawanabe, Motoaki},
  year = {2022},
  month = oct,
  number = {arXiv:2206.01323},
  eprint = {2206.01323},
  primaryclass = {cs, eess},
  publisher = {arXiv},
  doi = {10.48550/arXiv.2206.01323},
  urldate = {2022-11-18},
  archiveprefix = {arxiv}
}

@article{koMultiScaleNeuralNetwork2021,
  title = {Multi-{{Scale Neural Network}} for {{EEG Representation Learning}} in {{BCI}}},
  author = {Ko, Wonjun and Jeon, Eunjin and Jeong, Seungwoo and Suk, Heung-Il},
  year = {2021},
  month = may,
  journal = {IEEE Computational Intelligence Magazine},
  volume = {16},
  number = {2},
  pages = {31--45},
  issn = {1556-603X, 1556-6048},
  doi = {10.1109/MCI.2021.3061875},
  urldate = {2022-09-19}
}

@article{kostasBENDRUsingTransformers2021,
  title = {{{BENDR}}: {{Using Transformers}} and a {{Contrastive Self-Supervised Learning Task}} to {{Learn From Massive Amounts}} of {{EEG Data}}},
  shorttitle = {{{BENDR}}},
  author = {Kostas, Demetres and {Aroca-Ouellette}, St{\'e}phane and Rudzicz, Frank},
  year = {2021},
  month = jun,
  journal = {Frontiers in Human Neuroscience},
  volume = {15},
  pages = {653659},
  issn = {1662-5161},
  doi = {10.3389/fnhum.2021.653659},
  urldate = {2022-05-04},
  langid = {english}
}

@article{kullbackInformationSufficiency1951,
  title = {On {{Information}} and {{Sufficiency}}},
  author = {Kullback, S. and Leibler, R. A.},
  year = {1951},
  month = mar,
  journal = {The Annals of Mathematical Statistics},
  volume = {22},
  number = {1},
  pages = {79--86},
  publisher = {Institute of Mathematical Statistics},
  issn = {0003-4851, 2168-8990},
  doi = {10.1214/aoms/1177729694},
  urldate = {2023-05-04}
}

@article{kumaraguruTrustAwareRouting2021,
  type = {Article},
  title = {Trust Aware Routing Using Sunflower Sine Cosine-Based Stacked Autoencoder Approach for {{EEG}} Signal Classification in {{WSN}}},
  author = {Kumaraguru, Shanthi and Jebarani, M. R. Ebenezar},
  year = {2021},
  journal = {Journal Of High Speed Networks},
  volume = {27},
  number = {2},
  pages = {101--119},
  issn = {0926-6801},
  doi = {10.3233/jhs-210654}
}

@article{leeMotorImageryClassification2022,
  type = {Journal Article},
  title = {Motor Imagery Classification Using Inter-Task Transfer Learning via a Channel-Wise Variational Autoencoder-Based Convolutional Neural Network.},
  author = {Lee, Do-Yeun and Jeong, Ji-Hoon and Lee, Byeong-Hoo and Lee, Seong-Whan},
  year = {2022},
  journal = {IEEE transactions on neural systems and rehabilitation engineering},
  volume = {30},
  pages = {226--237},
  issn = {1558-0210},
  doi = {10.1109/tnsre.2022.3143836}
}

@article{liLatentFactorDecoding2020,
  type = {Journal Article},
  title = {Latent Factor Decoding of Multi-Channel {{EEG}} for Emotion Recognition through Autoencoder-like Neural Networks.},
  author = {Li, Xiang and Zhao, Zhigang and Song, Dawei and Zhang, Yazhou and Pan, Jingshan and Wu, Lu and Huo, Jidong and Niu, Chunyang and Wang, Di},
  year = {2020},
  journal = {Frontiers in neuroscience},
  volume = {14},
  pages = {87},
  issn = {1662-4548},
  doi = {10.3389/fnins.2020.00087}
}

@article{liMultitaskCollaborativeNetwork2024,
  type = {Journal Article},
  title = {Multi-Task Collaborative Network: {{Bridge}} the Supervised and Self-Supervised Learning for {{EEG}} Classification in {{RSVP}} Tasks},
  author = {Li, Hongxin and Tang, Jingsheng and Li, Wenqi and Dai, Wei and Liu, Yaru and Zhou, Zongtan},
  year = {2024},
  journal = {IEEE transactions on neural systems and rehabilitation engineering},
  volume = {32},
  pages = {638--651},
  issn = {1558-0210},
  doi = {10.1109/TNSRE.2024.3357863}
}

@article{linClassificationEpilepticEEG2016,
  type = {Proceedings Paper},
  title = {Classification of Epileptic {{EEG}} Signals with Stacked Sparse Autoencoder Based on Deep Learning},
  author = {Lin, Qin and Ye, Shu-qun and Huang, Xiu-mei and Li, Si-you and Zhang, Mei-zhen and Xue, Yun and Chen, Wen-Sheng},
  year = {2016},
  journal = {Intelligent Computing Methodologies},
  volume = {9773},
  pages = {802--810},
  issn = {0302-9743},
  doi = {10.1007/978-3-319-42297-8_74}
}

@article{lindermanFastInterpolationbasedTSNE2019,
  title = {Fast Interpolation-Based t-{{SNE}} for Improved Visualization of Single-Cell {{RNA-seq}} Data},
  author = {Linderman, George C. and Rachh, Manas and Hoskins, Jeremy G. and Steinerberger, Stefan and Kluger, Yuval},
  year = {2019},
  month = mar,
  journal = {Nature Methods},
  volume = {16},
  number = {3},
  pages = {243--245},
  publisher = {Nature Publishing Group},
  issn = {1548-7105},
  doi = {10.1038/s41592-018-0308-4},
  urldate = {2023-03-30},
  copyright = {2019 The Author(s), under exclusive licence to Springer Nature America, Inc.},
  langid = {english}
}

@inproceedings{linMicrosoftCOCOCommon2014,
  title = {Microsoft {{COCO}}: {{Common Objects}} in {{Context}}},
  shorttitle = {Microsoft {{COCO}}},
  booktitle = {Computer {{Vision}} -- {{ECCV}} 2014},
  author = {Lin, Tsung-Yi and Maire, Michael and Belongie, Serge and Hays, James and Perona, Pietro and Ramanan, Deva and Doll{\'a}r, Piotr and Zitnick, C. Lawrence},
  editor = {Fleet, David and Pajdla, Tomas and Schiele, Bernt and Tuytelaars, Tinne},
  year = {2014},
  series = {Lecture {{Notes}} in {{Computer Science}}},
  pages = {740--755},
  publisher = {Springer International Publishing},
  address = {Cham},
  doi = {10.1007/978-3-319-10602-1_48},
  isbn = {978-3-319-10602-1},
  langid = {english}
}

@article{liuDistinguishableSpatialspectralFeature2021,
  type = {Journal Article},
  title = {Distinguishable Spatial-Spectral Feature Learning Neural Network Framework for Motor Imagery-Based Brain-Computer Interface.},
  author = {Liu, Chang and Jin, Jing and Xu, Ren and Li, Shurui and Zuo, Cili and Sun, Hao and Wang, Xingyu and Cichocki, Andrzej},
  year = {2021},
  journal = {Journal of neural engineering},
  volume = {18},
  number = {4},
  issn = {1741-2552},
  doi = {10.1088/1741-2552/ac1d36}
}

@article{liuEEGBasedEmotionClassification2020,
  type = {Journal Article},
  title = {{{EEG-Based}} Emotion Classification Using a Deep Neural Network and Sparse Autoencoder.},
  author = {Liu, Junxiu and Wu, Guopei and Luo, Yuling and Qiu, Senhui and Yang, Su and Li, Wei and Bi, Yifei},
  year = {2020},
  journal = {Frontiers in systems neuroscience},
  volume = {14},
  pages = {43},
  issn = {1662-5137},
  doi = {10.3389/fnsys.2020.00043}
}

@article{liVariationalAutoencoderBased2019,
  type = {Proceedings Paper},
  title = {Variational Autoencoder Based Latent Factor Decoding of Multichannel {{EEG}} for Emotion Recognition},
  author = {Li, Xiang and Zhao, Zhigang and Song, Dawei and Zhang, Yazhou and Niu, Chunyang and Zhang, Junwei and Huo, Jidong and Li, Jing},
  year = {2019},
  journal = {IEEE International Conference On Bioinformatics And Biomedicine},
  pages = {684--687},
  issn = {2156-1125},
  doi = {10.1109/bibm47256.2019.8983341}
}

@misc{liVisualBERTSimplePerformant2019,
  title = {{{VisualBERT}}: {{A Simple}} and {{Performant Baseline}} for {{Vision}} and {{Language}}},
  shorttitle = {{{VisualBERT}}},
  author = {Li, Liunian Harold and Yatskar, Mark and Yin, Da and Hsieh, Cho-Jui and Chang, Kai-Wei},
  year = {2019},
  publisher = {arXiv},
  doi = {10.48550/ARXIV.1908.03557},
  urldate = {2023-03-21},
  copyright = {arXiv.org perpetual, non-exclusive license}
}

@article{maatenVisualizingDataUsing2008,
  title = {Visualizing {{Data}} Using T-{{SNE}}},
  author = {van der Maaten, Laurens and Hinton, Geoffrey},
  year = {2008},
  journal = {Journal of Machine Learning Research},
  volume = {9},
  number = {86},
  pages = {2579--2605},
  issn = {1533-7928},
  urldate = {2023-03-28}
}

@article{mammoneAutoEncoderFilterBank2023,
  type = {Journal Article},
  title = {{{AutoEncoder}} Filter Bank Common Spatial Patterns to Decode Motor Imagery from {{EEG}}},
  author = {Mammone, Nadia and Ieracitano, Cosimo and Adeli, Hojjat and Morabito, Francesco C},
  year = {2023},
  journal = {IEEE journal of biomedical and health informatics},
  volume = {27},
  number = {5},
  pages = {2365--2376},
  issn = {2168-2208},
  doi = {10.1109/JBHI.2023.3243698}
}

@misc{mcinnesPerformanceComparisonDimension,
  title = {Performance {{Comparison}} of {{Dimension Reduction Implementations}}. {{UMAP}} Documentation},
  author = {McInnes, Leland},
  journal = {UMAP documentation},
  urldate = {2024-05-17},
  howpublished = {https://umap-learn.readthedocs.io/en/latest/benchmarking.html}
}

@misc{mcinnesUMAPUniformManifold2020,
  title = {{{UMAP}}: {{Uniform Manifold Approximation}} and {{Projection}} for {{Dimension Reduction}}},
  shorttitle = {{{UMAP}}},
  author = {McInnes, Leland and Healy, John and Melville, James},
  year = {2020},
  month = sep,
  number = {arXiv:1802.03426},
  eprint = {1802.03426},
  primaryclass = {cs, stat},
  publisher = {arXiv},
  doi = {10.48550/arXiv.1802.03426},
  urldate = {2023-03-28},
  archiveprefix = {arxiv}
}

@article{mirzaeiEEGMotorImagery2021,
  type = {Article},
  title = {{{EEG}} Motor Imagery Classification Using Dynamic Connectivity Patterns and Convolutional Autoencoder},
  author = {Mirzaei, Sayeh and Ghasemi, Parisa},
  year = {2021},
  journal = {Biomedical Signal Processing And Control},
  volume = {68},
  issn = {1746-8094},
  doi = {10.1016/j.bspc.2021.102584}
}

@inproceedings{mohsenvandContrastiveRepresentationLearning2020,
  title = {Contrastive {{Representation Learning}} for {{Electroencephalogram Classification}}},
  booktitle = {Proceedings of the {{Machine Learning}} for {{Health NeurIPS Workshop}}},
  author = {Mohsenvand, Mostafa Neo and Izadi, Mohammad Rasool and Maes, Pattie},
  year = {2020},
  month = nov,
  pages = {238--253},
  publisher = {PMLR},
  issn = {2640-3498},
  urldate = {2023-10-03},
  langid = {english}
}

@inproceedings{nairImprovedApproachEEG2018,
  type = {Proceedings Paper},
  title = {An Improved Approach for {{EEG}} Signal Classification Using Autoencoder},
  booktitle = {8th {{International Symposium On Embedded Computing And System Design}}},
  author = {Nair, Abhijith V. and Kumar, Kodidasu Murali and Mathew, Jimson},
  year = {2018},
  pages = {6--10},
  doi = {10.1109/ised.2018.8704011}
}

@article{nejedlyUtilizationTemporalAutoencoder2023,
  type = {Journal Article},
  title = {Utilization of Temporal Autoencoder for Semi-Supervised Intracranial {{EEG}} Clustering and Classification},
  author = {Nejedly, Petr and Kremen, Vaclav and Lepkova, Kamila and Mivalt, Filip and Sladky, Vladimir and Pridalova, Tereza and Plesinger, Filip and Jurak, Pavel and Pail, Martin and Brazdil, Milan and Klimes, Petr and Worrell, Gregory},
  year = {2023},
  journal = {Scientific reports},
  volume = {13},
  number = {1},
  pages = {744},
  issn = {2045-2322},
  doi = {10.1038/s41598-023-27978-6}
}

@article{obeidTempleUniversityHospital2016,
  title = {The {{Temple University Hospital EEG Data Corpus}}},
  author = {Obeid, Iyad and Picone, Joseph},
  year = {2016},
  month = may,
  journal = {Frontiers in Neuroscience},
  volume = {10},
  issn = {1662-453X},
  doi = {10.3389/fnins.2016.00196},
  urldate = {2023-03-17}
}

@misc{oskolkovTSNEVsUMAP2020,
  title = {{{tSNE}} vs. {{UMAP}}: {{Global Structure}}},
  author = {Oskolkov, Nikolay},
  year = {2020},
  month = mar,
  journal = {Medium},
  urldate = {2024-05-17},
  howpublished = {https://towardsdatascience.com/tsne-vs-umap-global-structure-4d8045acba17}
}

@article{ouImprovedSelfsupervisedLearning2022,
  title = {An Improved Self-Supervised Learning for {{EEG}} Classification},
  author = {Ou, Yanghan and Sun, Siqin and Gan, Haitao and Zhou, Ran and Yang, Zhi},
  year = {2022},
  journal = {Mathematical Biosciences and Engineering},
  volume = {19},
  number = {7},
  pages = {6907--6922},
  issn = {1551-0018},
  doi = {10.3934/mbe.2022325},
  urldate = {2022-11-11}
}

@article{ozdenizciLearningInvariantRepresentations2020,
  title = {Learning {{Invariant Representations From EEG}} via {{Adversarial Inference}}},
  author = {{\"O}zdenizci, Ozan and Wang, Ye and {Koike-Akino}, Toshiaki and Erdo{\u g}mu{\c s}, Deniz},
  year = {2020},
  journal = {IEEE Access},
  volume = {8},
  pages = {27074--27085},
  issn = {2169-3536},
  doi = {10.1109/ACCESS.2020.2971600}
}

@inproceedings{ozdenizciTransferLearningBrainComputer2019,
  title = {Transfer {{Learning}} in {{Brain-Computer Interfaces}} with {{Adversarial Variational Autoencoders}}},
  booktitle = {2019 9th {{International IEEE}}/{{EMBS Conference}} on {{Neural Engineering}} ({{NER}})},
  author = {{\"O}zdenizci, Ozan and Wang, Ye and {Koike-Akino}, Toshiaki and Erdogmus, Deniz},
  year = {2019},
  month = mar,
  pages = {207--210},
  publisher = {IEEE},
  address = {San Francisco, CA, USA},
  doi = {10.1109/NER.2019.8716897},
  urldate = {2022-12-20},
  isbn = {978-1-5386-7921-0}
}

@article{panSurveyTransferLearning2010,
  title = {A {{Survey}} on {{Transfer Learning}}},
  author = {Pan, Sinno Jialin and Yang, Qiang},
  year = {2010},
  month = oct,
  journal = {IEEE Transactions on Knowledge and Data Engineering},
  volume = {22},
  number = {10},
  pages = {1345--1359},
  issn = {1558-2191},
  doi = {10.1109/TKDE.2009.191}
}

@inproceedings{parashivaNewChannelSelection2019,
  title = {A {{New Channel Selection Method}} Using {{Autoencoder}} for {{Motor Imagery}} Based {{Brain Computer Interface}}},
  booktitle = {2019 {{IEEE International Conference}} on {{Systems}}, {{Man}} and {{Cybernetics}} ({{SMC}})},
  author = {Parashiva, Praveen K. and Vinod, A. P.},
  year = {2019},
  month = oct,
  pages = {3641--3646},
  publisher = {IEEE},
  address = {Bari, Italy},
  doi = {10.1109/SMC.2019.8914251},
  urldate = {2022-11-11},
  isbn = {978-1-72814-569-3}
}

@article{parijaAutoencoderbasedImprovedDeep2022,
  type = {Article},
  title = {Autoencoder-Based Improved Deep Learning Approach for Schizophrenic {{EEG}} Signal Classification},
  author = {Parija, Sebamai and Sahani, Mrutyunjaya and Bisoi, Ranjeeta and Dash, P. K.},
  year = {2022},
  journal = {Pattern Analysis And Applications},
  issn = {1433-7541},
  doi = {10.1007/s10044-022-01107-x}
}

@article{peiDecodingAsynchronousReaching2018,
  type = {Article},
  title = {Decoding Asynchronous Reaching in Electroencephalography Using Stacked Autoencoders},
  author = {Pei, Dingyi and Burns, Martin and Chandramouli, Rajarathnam and Vinjamuri, Ramana},
  year = {2018},
  journal = {IEEE access : practical innovations, open solutions},
  volume = {6},
  pages = {52889--52898},
  issn = {2169-3536},
  doi = {10.1109/access.2018.2869687}
}

@inproceedings{petrutiuEnhancingClassificationEEG2020,
  title = {Enhancing the {{Classification}} of {{EEG Signals}} Using {{Wasserstein Generative Adversarial Networks}}},
  booktitle = {2020 {{IEEE}} 16th {{International Conference}} on {{Intelligent Computer Communication}} and {{Processing}} ({{ICCP}})},
  author = {Petrutiu, Vlad Mihai and Palcu, Liana Daniela and Lemnaru, Camelia and Dinsoreanu, Mihaela and Potolea, Rodica and Mursesan, Raul and Moca, Vlad Vasile},
  year = {2020},
  month = sep,
  pages = {29--34},
  publisher = {IEEE},
  address = {Cluj-Napoca, Romania},
  doi = {10.1109/ICCP51029.2020.9266157},
  urldate = {2022-11-11},
  isbn = {978-1-72819-080-8}
}

@article{phadikarUnsupervisedFeatureExtraction2023,
  type = {Article},
  title = {Unsupervised Feature Extraction with Autoencoders for {{EEG}} Based Multiclass Motor Imagery {{BCI}}},
  author = {Phadikar, Souvik and Sinha, Nidul and Ghosh, Rajdeep},
  year = {2023},
  journal = {Expert Systems With Applications},
  volume = {213},
  issn = {0957-4174},
  doi = {10.1016/j.eswa.2022.118901}
}

@article{phunruangsakaoMultibranchConvolutionalNeural2022,
  type = {Journal Article},
  title = {Multibranch Convolutional Neural Network with Contrastive Representation Learning for Decoding Same Limb Motor Imagery Tasks},
  author = {Phunruangsakao, Chatrin and Achanccaray, David and Izumi, Shin-Ichi and Hayashibe, Mitsuhiro},
  year = {2022},
  journal = {Frontiers in human neuroscience},
  volume = {16},
  pages = {1032724},
  issn = {1662-5161},
  doi = {10.3389/fnhum.2022.1032724}
}

@article{prabhakarSASDLRBATQSparse2022,
  type = {Journal Article},
  title = {{{SASDL}} and {{RBATQ}}: {{Sparse}} Autoencoder with Swarm Based Deep Learning and Reinforcement Based {{Q-learning}} for {{EEG}} Classification.},
  author = {Prabhakar, Sunil Kumar and Lee, Seong-Whan},
  year = {2022},
  journal = {IEEE open journal of engineering in medicine and biology},
  volume = {3},
  pages = {58--68},
  issn = {2644-1276},
  doi = {10.1109/ojemb.2022.3161837}
}

@article{qiuDenoisingSparseAutoencoderbased2018,
  type = {Journal Article},
  title = {Denoising Sparse Autoencoder-Based Ictal {{EEG}} Classification.},
  author = {Qiu, Yang and Zhou, Weidong and Yu, Nana and Du, Peidong},
  year = {2018},
  journal = {IEEE transactions on neural systems and rehabilitation engineering},
  volume = {26},
  number = {9},
  pages = {1717--1726},
  issn = {1558-0210},
  doi = {10.1109/tnsre.2018.2864306}
}

@misc{radfordUnsupervisedRepresentationLearning2016,
  title = {Unsupervised {{Representation Learning}} with {{Deep Convolutional Generative Adversarial Networks}}},
  author = {Radford, Alec and Metz, Luke and Chintala, Soumith},
  year = {2016},
  month = jan,
  number = {arXiv:1511.06434},
  eprint = {1511.06434},
  primaryclass = {cs},
  publisher = {arXiv},
  doi = {10.48550/arXiv.1511.06434},
  urldate = {2022-12-14},
  archiveprefix = {arxiv}
}

@misc{rajpurkarSQuAD1000002016,
  title = {{{SQuAD}}: 100,000+ {{Questions}} for {{Machine Comprehension}} of {{Text}}},
  shorttitle = {{{SQuAD}}},
  author = {Rajpurkar, Pranav and Zhang, Jian and Lopyrev, Konstantin and Liang, Percy},
  year = {2016},
  month = oct,
  number = {arXiv:1606.05250},
  eprint = {1606.05250},
  primaryclass = {cs},
  publisher = {arXiv},
  doi = {10.48550/arXiv.1606.05250},
  urldate = {2023-04-20},
  archiveprefix = {arxiv}
}

@article{ranHybridAutoencoderFramework2022,
  type = {Journal Article},
  title = {A Hybrid Autoencoder Framework of Dimensionality Reduction for Brain-Computer Interface Decoding.},
  author = {Ran, Xingchen and Chen, Weidong and Yvert, Blaise and Zhang, Shaomin},
  year = {2022},
  journal = {Computers in biology and medicine},
  volume = {148},
  pages = {105871},
  issn = {1879-0534},
  doi = {10.1016/j.compbiomed.2022.105871}
}

@misc{rommelCADDAClasswiseAutomatic2022,
  title = {{{CADDA}}: {{Class-wise Automatic Differentiable Data Augmentation}} for {{EEG Signals}}},
  shorttitle = {{{CADDA}}},
  author = {Rommel, C{\'e}dric and Moreau, Thomas and Paillard, Joseph and Gramfort, Alexandre},
  year = {2022},
  month = feb,
  number = {arXiv:2106.13695},
  eprint = {2106.13695},
  primaryclass = {cs},
  institution = {arXiv},
  doi = {10.48550/arXiv.2106.13695},
  urldate = {2022-05-24},
  archiveprefix = {arxiv}
}

@article{rommelDataAugmentationLearning2022,
  title = {Data Augmentation for Learning Predictive Models on {{EEG}}: A Systematic Comparison},
  shorttitle = {Data Augmentation for Learning Predictive Models on {{EEG}}},
  author = {Rommel, C{\'e}dric and Paillard, Joseph and Moreau, Thomas and Gramfort, Alexandre},
  year = {2022},
  journal = {Journal of Neural Engineering},
  issn = {1741-2552},
  doi = {10.1088/1741-2552/aca220},
  urldate = {2022-11-18},
  langid = {english}
}

@misc{rommelDeepInvariantNetworks2022,
  title = {Deep Invariant Networks with Differentiable Augmentation Layers},
  author = {Rommel, C{\'e}dric and Moreau, Thomas and Gramfort, Alexandre},
  year = {2022},
  month = oct,
  number = {arXiv:2202.02142},
  eprint = {2202.02142},
  primaryclass = {cs},
  publisher = {arXiv},
  doi = {10.48550/arXiv.2202.02142},
  urldate = {2022-11-18},
  archiveprefix = {arxiv}
}

@article{royDeepLearningbasedElectroencephalography2019,
  title = {Deep Learning-Based Electroencephalography Analysis: A Systematic Review},
  shorttitle = {Deep Learning-Based Electroencephalography Analysis},
  author = {Roy, Yannick and Banville, Hubert and Albuquerque, Isabela and Gramfort, Alexandre and Falk, Tiago H and Faubert, Jocelyn},
  year = {2019},
  month = oct,
  journal = {Journal of Neural Engineering},
  volume = {16},
  number = {5},
  pages = {051001},
  issn = {1741-2560, 1741-2552},
  doi = {10.1088/1741-2552/ab260c},
  urldate = {2022-05-04},
  langid = {english}
}

@article{royMIEEGGANGeneratingArtificial2020,
  type = {Proceedings Paper},
  title = {{{MIEEG-GAN}}: {{Generating}} Artificial Motor Imagery Electroencephalography Signals},
  author = {Roy, Sujit and Dora, Shirin and McCreadie, Karl and Prasad, Girijesh},
  year = {2020},
  journal = {International Joint Conference On Neural Networks},
  issn = {2161-4393},
  doi = {10.1109/ijcnn48605.2020.9206942}
}

@misc{schneiderLearnableLatentEmbeddings2022,
  title = {Learnable Latent Embeddings for Joint Behavioral and Neural Analysis},
  author = {Schneider, Steffen and Lee, Jin Hwa and Mathis, Mackenzie Weygandt},
  year = {2022},
  month = oct,
  number = {arXiv:2204.00673},
  eprint = {2204.00673},
  primaryclass = {cs, q-bio},
  publisher = {arXiv},
  doi = {10.48550/arXiv.2204.00673},
  urldate = {2023-04-14},
  archiveprefix = {arxiv}
}

@inproceedings{schroffFaceNetUnifiedEmbedding2015,
  title = {{{FaceNet}}: {{A}} Unified Embedding for Face Recognition and Clustering},
  shorttitle = {{{FaceNet}}},
  booktitle = {2015 {{IEEE Conference}} on {{Computer Vision}} and {{Pattern Recognition}} ({{CVPR}})},
  author = {Schroff, Florian and Kalenichenko, Dmitry and Philbin, James},
  year = {2015},
  month = jun,
  pages = {815--823},
  publisher = {IEEE},
  address = {Boston, MA, USA},
  doi = {10.1109/CVPR.2015.7298682},
  urldate = {2022-05-03},
  isbn = {978-1-4673-6964-0},
  langid = {english}
}

@misc{serdyukInvariantRepresentationsNoisy2016,
  title = {Invariant {{Representations}} for {{Noisy Speech Recognition}}},
  author = {Serdyuk, Dmitriy and Audhkhasi, Kartik and Brakel, Phil{\'e}mon and Ramabhadran, Bhuvana and Thomas, Samuel and Bengio, Yoshua},
  year = {2016},
  month = nov,
  number = {arXiv:1612.01928},
  eprint = {1612.01928},
  primaryclass = {cs, stat},
  publisher = {arXiv},
  doi = {10.48550/arXiv.1612.01928},
  urldate = {2022-12-20},
  archiveprefix = {arxiv}
}

@article{songEEGConformerConvolutional2023,
  title = {{{EEG Conformer}}: {{Convolutional Transformer}} for {{EEG Decoding}} and {{Visualization}}},
  shorttitle = {{{EEG Conformer}}},
  author = {Song, Yonghao and Zheng, Qingqing and Liu, Bingchuan and Gao, Xiaorong},
  year = {2023},
  journal = {IEEE Transactions on Neural Systems and Rehabilitation Engineering},
  volume = {31},
  pages = {710--719},
  issn = {1558-0210},
  doi = {10.1109/TNSRE.2022.3230250},
  urldate = {2023-11-15}
}

@article{sosulskiImprovingCovarianceMatrices2021,
  title = {Improving {{Covariance Matrices Derived}} from {{Tiny Training Datasets}} for the {{Classification}} of {{Event-Related Potentials}} with {{Linear Discriminant Analysis}}},
  author = {Sosulski, Jan and Kemmer, Jan-Philipp and Tangermann, Michael},
  year = {2021},
  month = jul,
  journal = {Neuroinformatics},
  volume = {19},
  number = {3},
  pages = {461--476},
  issn = {1559-0089},
  doi = {10.1007/s12021-020-09501-8},
  urldate = {2024-04-11},
  langid = {english}
}

@article{sosulskiIntroducingBlockToeplitzCovariance2022,
  title = {Introducing Block-{{Toeplitz}} Covariance Matrices to Remaster Linear Discriminant Analysis for Event-Related Potential Brain--Computer Interfaces},
  author = {Sosulski, Jan and Tangermann, Michael},
  year = {2022},
  month = nov,
  journal = {Journal of Neural Engineering},
  volume = {19},
  number = {6},
  pages = {066001},
  publisher = {IOP Publishing},
  issn = {1741-2552},
  doi = {10.1088/1741-2552/ac9c98},
  urldate = {2023-05-03},
  langid = {english}
}

@misc{sosulskiUMMUnsupervisedMeandifference2023,
  title = {{{UMM}}: {{Unsupervised Mean-difference Maximization}}},
  shorttitle = {{{UMM}}},
  author = {Sosulski, Jan and Tangermann, Michael},
  year = {2023},
  month = jun,
  number = {arXiv:2306.11830},
  eprint = {2306.11830},
  primaryclass = {cs, stat},
  publisher = {arXiv},
  doi = {10.48550/arXiv.2306.11830},
  urldate = {2023-08-07},
  archiveprefix = {arxiv}
}

@article{stepheMotorImageryEEG2022,
  type = {Article},
  title = {Motor Imagery {{EEG}} Recognition Using Deep Generative Adversarial Network with {{EMD}} for {{BCI}} Applications},
  author = {Stephe, Stephan and Jayasankar, Thangaiyan and Kumar, Kalimuthu Vinoth},
  year = {2022},
  journal = {Tehnicki Vjesnik-Technical Gazette},
  volume = {29},
  number = {1},
  pages = {92--100},
  issn = {1330-3651},
  doi = {10.17559/tv-20210121112228}
}

@article{tanAutoencoderbasedTransferLearning2019,
  type = {Article},
  title = {Autoencoder-Based Transfer Learning in Brain-Computer Interface for Rehabilitation Robot},
  author = {Tan, Chuanqi and Sun, Fuchun and Fang, Bin and Kong, Tao and Zhang, Wenchang},
  year = {2019},
  journal = {International Journal Of Advanced Robotic Systems},
  volume = {16},
  number = {2},
  issn = {1729-8814},
  doi = {10.1177/1729881419840860}
}

@article{tangMotorImageryEEG2019,
  type = {Journal Article},
  title = {Motor Imagery {{EEG}} Recognition with {{KNN-based}} Smooth Auto-Encoder.},
  author = {Tang, Xianlun and Wang, Ting and Du, Yiming and Dai, Yuyan},
  year = {2019},
  journal = {Artificial intelligence in medicine},
  volume = {101},
  pages = {101747},
  issn = {1873-2860},
  doi = {10.1016/j.artmed.2019.101747}
}

@article{thielenFullCalibrationZero2021,
  title = {From Full Calibration to Zero Training for a Code-Modulated Visual Evoked Potentials for Brain--Computer Interface},
  author = {Thielen, J. and Marsman, P. and Farquhar, J. and Desain, P.},
  year = {2021},
  month = apr,
  journal = {Journal of Neural Engineering},
  volume = {18},
  number = {5},
  pages = {056007},
  publisher = {IOP Publishing},
  issn = {1741-2552},
  doi = {10.1088/1741-2552/abecef},
  urldate = {2023-08-18},
  langid = {english}
}

@article{tippingProbabilisticPrincipalComponent1999,
  title = {Probabilistic {{Principal Component Analysis}}},
  author = {Tipping, Michael E. and Bishop, Christopher M.},
  year = {1999},
  month = sep,
  journal = {Journal of the Royal Statistical Society Series B: Statistical Methodology},
  volume = {61},
  number = {3},
  pages = {611--622},
  issn = {1369-7412, 1467-9868},
  doi = {10.1111/1467-9868.00196},
  urldate = {2023-03-27},
  langid = {english}
}

@article{tomonagaExperimentsClassificationElectroencephalography2017,
  type = {Article},
  title = {Experiments on Classification of Electroencephalography ({{EEG}}) Signals in Imagination of Direction Using Stacked Autoencoder},
  author = {Tomonaga, Kenta and Hayakawa, Takuya and Kobayashi, Jun},
  year = {2017},
  journal = {Journal Of Robotics Networking And Artificial Life},
  volume = {4},
  number = {2},
  pages = {124--128},
  issn = {2352-6386},
  doi = {10.2991/jrnal.2017.4.2.4}
}

@misc{tormaEEGWaveDenoisingDiffusion2023,
  title = {{{EEGWave}}: A {{Denoising Diffusion Probablistic Approach}} for {{EEG Signal Generation}}},
  author = {Torma, Szabolcs and Szegletes, Dr Luca},
  year = {2023},
  number = {10275},
  eprint = {10275},
  publisher = {EasyChair},
  archiveprefix = {EasyChair},
  langid = {english}
}

@article{vanheckeZotero2008,
  title = {Zotero},
  author = {Vanhecke, Thomas E.},
  year = {2008},
  month = jul,
  journal = {Journal of the Medical Library Association : JMLA},
  volume = {96},
  number = {3},
  pages = {275--276},
  issn = {1536-5050},
  doi = {10.3163/1536-5050.96.3.022},
  urldate = {2023-07-10},
  pmcid = {PMC2479046},
  pmid = {null}
}

@inproceedings{vanhornINaturalistSpeciesClassification2018,
  title = {The {{INaturalist Species Classification}} and {{Detection Dataset}}},
  booktitle = {Proceedings of the {{IEEE Conference}} on {{Computer Vision}} and {{Pattern Recognition}}},
  author = {Van Horn, Grant and Mac Aodha, Oisin and Song, Yang and Cui, Yin and Sun, Chen and Shepard, Alex and Adam, Hartwig and Perona, Pietro and Belongie, Serge},
  year = {2018},
  pages = {8769--8778},
  urldate = {2023-04-20}
}

@inproceedings{vaswaniAttentionAllYou2017,
  title = {Attention Is {{All}} You {{Need}}},
  booktitle = {Advances in {{Neural Information Processing Systems}}},
  author = {Vaswani, Ashish and Shazeer, Noam and Parmar, Niki and Uszkoreit, Jakob and Jones, Llion and Gomez, Aidan N and Kaiser, {\L}ukasz and Polosukhin, Illia},
  year = {2017},
  volume = {30},
  publisher = {Curran Associates, Inc.},
  urldate = {2022-07-11}
}

@inproceedings{vondrickGeneratingVideosScene2016,
  title = {Generating {{Videos}} with {{Scene Dynamics}}},
  booktitle = {Advances in {{Neural Information Processing Systems}}},
  author = {Vondrick, Carl and Pirsiavash, Hamed and Torralba, Antonio},
  year = {2016},
  volume = {29},
  publisher = {Curran Associates, Inc.},
  urldate = {2022-12-16}
}

@misc{wangGLUEMultiTaskBenchmark2019,
  title = {{{GLUE}}: {{A Multi-Task Benchmark}} and {{Analysis Platform}} for {{Natural Language Understanding}}},
  shorttitle = {{{GLUE}}},
  author = {Wang, Alex and Singh, Amanpreet and Michael, Julian and Hill, Felix and Levy, Omer and Bowman, Samuel R.},
  year = {2019},
  month = feb,
  number = {arXiv:1804.07461},
  eprint = {1804.07461},
  primaryclass = {cs},
  publisher = {arXiv},
  urldate = {2023-04-20},
  archiveprefix = {arxiv}
}

@article{wangNovelAlgorithmicStructure2023,
  type = {Journal Article},
  title = {A Novel Algorithmic Structure of {{EEG}} Channel Attention Combined with Swin Transformer for Motor Patterns Classification},
  author = {Wang, Han and Cao, Lei and Huang, Chenxi and Jia, Jie and Dong, Yilin and Fan, Chunjiang and {de Albuquerque}, Victor Hugo C},
  year = {2023},
  journal = {IEEE transactions on neural systems and rehabilitation engineering},
  volume = {31},
  pages = {3132--3141},
  issn = {1558-0210},
  doi = {10.1109/TNSRE.2023.3297654}
}

@inproceedings{wei2021BEETLCompetition2022,
  title = {2021 {{BEETL Competition}}: {{Advancing Transfer Learning}} for {{Subject Independence}} \& {{Heterogenous EEG Data Sets}}},
  shorttitle = {2021 {{BEETL Competition}}},
  booktitle = {Proceedings of the {{NeurIPS}} 2021 {{Competitions}} and {{Demonstrations Track}}},
  author = {Wei, Xiaoxi and Faisal, A. Aldo and {Grosse-Wentrup}, Moritz and Gramfort, Alexandre and Chevallier, Sylvain and Jayaram, Vinay and Jeunet, Camille and Bakas, Stylianos and Ludwig, Siegfried and Barmpas, Konstantinos and Bahri, Mehdi and Panagakis, Yannis and Laskaris, Nikolaos and Adamos, Dimitrios A. and Zafeiriou, Stefanos and Duong, William C. and Gordon, Stephen M. and Lawhern, Vernon J. and {\'S}liwowski, Maciej and Rouanne, Vincent and Tempczyk, Piotr},
  year = {2022},
  month = jul,
  pages = {205--219},
  publisher = {PMLR},
  issn = {2640-3498},
  urldate = {2023-02-13},
  langid = {english}
}

@article{wuTransferLearningEEGBased2022,
  title = {Transfer {{Learning}} for {{EEG-Based Brain}}--{{Computer Interfaces}}: {{A Review}} of {{Progress Made Since}} 2016},
  shorttitle = {Transfer {{Learning}} for {{EEG-Based Brain}}--{{Computer Interfaces}}},
  author = {Wu, Dongrui and Xu, Yifan and Lu, Bao-Liang},
  year = {2022},
  month = mar,
  journal = {IEEE Transactions on Cognitive and Developmental Systems},
  volume = {14},
  number = {1},
  pages = {4--19},
  issn = {2379-8939},
  doi = {10.1109/TCDS.2020.3007453},
  urldate = {2024-05-16}
}

@article{xieMotorImageryEEG2020,
  type = {Proceedings Paper},
  title = {Motor Imagery {{EEG}} Recognition Based on Scheduled Empirical Mode Decomposition and Adaptive Denoising Autoencoders},
  author = {Xie, Tao and Ma, Weichang and Li, Xingchen and Li, Wei and Hao, Bohui and Tang, Xianlun},
  year = {2020},
  journal = {IEEE Chinese Automation Congress},
  pages = {1528--1532},
  issn = {2688-092X},
  doi = {10.1109/cac51589.2020.9327855}
}

@article{xuRepresentationLearningMotor2021,
  title = {Representation {{Learning}} for {{Motor Imagery Recognition}} with {{Deep Neural Network}}},
  author = {Xu, Fangzhou and Rong, Fenqi and Miao, Yunjing and Sun, Yanan and Dong, Gege and Li, Han and Li, Jincheng and Wang, Yuandong and Leng, Jiancai},
  year = {2021},
  month = jan,
  journal = {Electronics},
  volume = {10},
  number = {2},
  pages = {112},
  issn = {2079-9292},
  doi = {10.3390/electronics10020112},
  urldate = {2022-09-19},
  langid = {english}
}

@article{yanEEGSignalClassification2016,
  type = {Proceedings Paper},
  title = {An {{EEG}} Signal Classification Method Based on Sparse Auto-Encoders and Support Vector Machine},
  author = {Yan, Bo and Wang, Yong and Li, Yuheng and Gong, Yejiang and Guan, Lu and Yu, Sheng},
  year = {2016},
  journal = {IEEE International Conference On Communications In China},
  issn = {2377-8644},
  doi = {10.1109/iccchina.2016.7636897}
}

@article{yangAssessingCognitiveMental2019,
  type = {Journal Article},
  title = {Assessing Cognitive Mental Workload via {{EEG}} Signals and an Ensemble Deep Learning Classifier Based on Denoising Autoencoders.},
  author = {Yang, Shuo and Yin, Zhong and Wang, Yagang and Zhang, Wei and Wang, Yongxiong and Zhang, Jianhua},
  year = {2019},
  journal = {Computers in biology and medicine},
  volume = {109},
  pages = {159--170},
  issn = {1879-0534},
  doi = {10.1016/j.compbiomed.2019.04.034}
}

@misc{yangBIOTCrossdataBiosignal2023,
  title = {{{BIOT}}: {{Cross-data Biosignal Learning}} in the {{Wild}}},
  shorttitle = {{{BIOT}}},
  author = {Yang, Chaoqi and Westover, M. Brandon and Sun, Jimeng},
  year = {2023},
  month = may,
  number = {arXiv:2305.10351},
  eprint = {2305.10351},
  primaryclass = {cs, eess},
  publisher = {arXiv},
  doi = {10.48550/arXiv.2305.10351},
  urldate = {2023-08-07},
  archiveprefix = {arxiv}
}

@article{yangNovelDeepLearning2020,
  title = {A {{Novel Deep Learning Scheme}} for {{Motor Imagery EEG Decoding Based}} on {{Spatial Representation Fusion}}},
  author = {Yang, Jun and Ma, Zhengmin and Wang, Jin and Fu, Yunfa},
  year = {2020},
  journal = {IEEE Access},
  volume = {8},
  pages = {202100--202110},
  issn = {2169-3536},
  doi = {10.1109/ACCESS.2020.3035347},
  urldate = {2022-09-19}
}

@misc{yangSelfsupervisedEEGRepresentation2023,
  title = {Self-Supervised {{EEG Representation Learning}} for {{Automatic Sleep Staging}}},
  author = {Yang, Chaoqi and Xiao, Danica and Westover, M. Brandon and Sun, Jimeng},
  year = {2023},
  month = feb,
  number = {arXiv:2110.15278},
  eprint = {2110.15278},
  primaryclass = {cs, eess},
  publisher = {arXiv},
  doi = {10.48550/arXiv.2110.15278},
  urldate = {2023-07-23},
  archiveprefix = {arxiv}
}

@article{yaoEmotionClassificationBased2024,
  type = {Journal Article},
  title = {Emotion Classification Based on Transformer and {{CNN}} for {{EEG}} Spatial-Temporal Feature Learning},
  author = {Yao, Xiuzhen and Li, Tianwen and Ding, Peng and Wang, Fan and Zhao, Lei and Gong, Anmin and Nan, Wenya and Fu, Yunfa},
  year = {2024},
  journal = {Brain sciences},
  volume = {14},
  number = {3},
  issn = {2076-3425},
  doi = {10.3390/brainsci14030268}
}

@article{yinPhysiologicalsignalbasedMentalWorkload2019,
  type = {Article},
  title = {Physiological-Signal-Based Mental Workload Estimation via Transfer Dynamical Autoencoders in a Deep Learning Framework},
  author = {Yin, Zhong and Zhao, Mengyuan and Zhang, Wei and Wang, Yongxiong and Wang, Yagang and Zhang, Jianhua},
  year = {2019},
  journal = {Neurocomputing},
  volume = {347},
  pages = {212--229},
  issn = {0925-2312},
  doi = {10.1016/j.neucom.2019.02.061}
}

@article{yuAdaptiveEEGFeature2021,
  type = {Journal Article},
  title = {An Adaptive {{EEG}} Feature Extraction Method Based on Stacked Denoising Autoencoder for Mental Fatigue Connectivity.},
  author = {Yu, Zhongliang and Li, Lili and Zhang, Wenwei and Lv, Hangyuan and Liu, Yun and Khalique, Umair},
  year = {2021},
  journal = {Neural plasticity},
  volume = {2021},
  pages = {3965385},
  issn = {1687-5443},
  doi = {10.1155/2021/3965385}
}

@misc{zhangBrain2ObjectPrintingYour2020,
  title = {{{Brain2Object}}: {{Printing Your Mind}} from {{Brain Signals}} with {{Spatial Correlation Embedding}}},
  shorttitle = {{{Brain2Object}}},
  author = {Zhang, Xiang and Yao, Lina and Huang, Chaoran and Kanhere, Salil S. and Zhang, Dalin and Zhang, Yu},
  year = {2020},
  month = jun,
  number = {arXiv:1810.02223},
  eprint = {1810.02223},
  primaryclass = {cs},
  institution = {arXiv},
  doi = {10.48550/arXiv.1810.02223},
  urldate = {2022-09-20},
  archiveprefix = {arxiv}
}

@inproceedings{zhangConvertingYourThoughts2018,
  title = {Converting {{Your Thoughts}} to {{Texts}}: {{Enabling Brain Typing}} via {{Deep Feature Learning}} of {{EEG Signals}}},
  shorttitle = {Converting {{Your Thoughts}} to {{Texts}}},
  booktitle = {2018 {{IEEE International Conference}} on {{Pervasive Computing}} and {{Communications}} ({{PerCom}})},
  author = {Zhang, Xiang and Yao, Lina and Sheng, Quan Z. and Kanhere, Salil S. and Gu, Tao and Zhang, Dalin},
  year = {2018},
  month = mar,
  pages = {1--10},
  issn = {2474-249X},
  doi = {10.1109/PERCOM.2018.8444575}
}

@misc{zhangMultitaskGenerativeAdversarial2020,
  title = {Multi-Task {{Generative Adversarial Learning}} on {{Geometrical Shape Reconstruction}} from {{EEG Brain Signals}}},
  author = {Zhang, Xiang and Chen, Xiaocong and Dong, Manqing and Liu, Huan and Ge, Chang and Yao, Lina},
  year = {2020},
  month = feb,
  number = {arXiv:1907.13351},
  eprint = {1907.13351},
  primaryclass = {cs, eess},
  institution = {arXiv},
  doi = {10.48550/arXiv.1907.13351},
  urldate = {2022-09-20},
  archiveprefix = {arxiv}
}

@article{zhangRealizingApplicationEEG2021,
  type = {Journal Article},
  title = {Realizing the Application of {{EEG}} Modeling in {{BCI}} Classification: {{Based}} on a Conditional {{GAN}} Converter.},
  author = {Zhang, Xiaodong and Lu, Zhufeng and Zhang, Teng and Li, Hanzhe and Wang, Yachun and Tao, Qing},
  year = {2021},
  journal = {Frontiers in neuroscience},
  volume = {15},
  pages = {727394},
  issn = {1662-4548},
  doi = {10.3389/fnins.2021.727394}
}

@article{zhangSpectralTemporalFeature2019,
  type = {Journal Article},
  title = {Spectral and Temporal Feature Learning with Two-Stream Neural Networks for Mental Workload Assessment.},
  author = {Zhang, Pengbo and Wang, Xue and Chen, Junfeng and You, Wei and Zhang, Weihang},
  year = {2019},
  journal = {IEEE transactions on neural systems and rehabilitation engineering},
  volume = {27},
  number = {6},
  pages = {1149--1159},
  issn = {1558-0210},
  doi = {10.1109/tnsre.2019.2913400}
}

@article{zhaoDeepCNNModel2022,
  type = {Article},
  title = {Deep {{CNN}} Model Based on Serial-Parallel Structure Optimization for Four-Class Motor Imagery {{EEG}} Classification},
  author = {Zhao, Xuefei and Liu, Dong and Ma, Li and Liu, Quan and Chen, Kun and Xie, Shane and Ai, Qingsong},
  year = {2022},
  journal = {Biomedical Signal Processing And Control},
  volume = {72},
  issn = {1746-8094},
  doi = {10.1016/j.bspc.2021.103338}
}

@inproceedings{zhouLearningDeepFeatures2014,
  title = {Learning {{Deep Features}} for {{Scene Recognition}} Using {{Places Database}}},
  booktitle = {Advances in {{Neural Information Processing Systems}}},
  author = {Zhou, Bolei and Lapedriza, Agata and Xiao, Jianxiong and Torralba, Antonio and Oliva, Aude},
  year = {2014},
  volume = {27},
  publisher = {Curran Associates, Inc.},
  urldate = {2023-04-20}
}

@misc{zlatovPhysiologyinformedDataAugmentation2022,
  title = {Towards Physiology-Informed Data Augmentation for {{EEG-based BCIs}}},
  author = {Zlatov, Oleksandr and Blankertz, Benjamin},
  year = {2022},
  month = mar,
  number = {arXiv:2203.14392},
  eprint = {2203.14392},
  primaryclass = {cs, eess},
  publisher = {arXiv},
  doi = {10.48550/arXiv.2203.14392},
  urldate = {2023-08-10},
  archiveprefix = {arxiv}
}

\end{document}